\documentclass[pra,superscriptaddress,showpacs,twocolumn,amsfonts,amsmath,amssymb,longbibliography]{revtex4-1}

\usepackage{graphicx}
\usepackage{bm}

\begin{document}

\title{Pressure profiles of nonuniform two-dimensional atomic Fermi gases}
\author{Kirill Martiyanov}\author{Tatiana Barmashova}\author{Vasiliy Makhalov}
\affiliation{Institute of Applied Physics, Russian Academy of Sciences, ul.~Ulyanova 46, Nizhniy Novgorod, Russia}
\author{Andrey Turlapov}\email{turlapov@appl.sci-nnov.ru}
\affiliation{Institute of Applied Physics, Russian Academy of Sciences, ul.~Ulyanova 46, Nizhniy Novgorod, Russia}
\affiliation{N. I. Lobachevsky State University of Nizhni Novgorod, Nizhniy Novgorod, 603950, Russia}

\date{\today}

\begin{abstract}
Spatial profiles of the pressure have been measured in atomic Fermi gases with primarily 2D kinematics.
The in-plane motion of the particles is confined by a gaussian-shape potential.
The two-component deeply-degenerate Fermi gases are prepared at different values of the \textit{s}-wave attraction. The pressure profile is found using the force-balance equation, from the measured density profile and the trapping potential.
The pressure is compared to zero-temperature models within the local density approximation. In the weakly-interacting regime, the pressure lies above a Landau Fermi-liquid theory and below the ideal-Fermi-gas model, whose prediction coincides with that of the Cooper-pair mean-field theory. The values closest to the data are provided by the approach where the mean-field of Cooper pairs is supplemented with fluctuations.
In the regime of strong interactions, in response to the increasing attraction, the pressure shifts below this model reaching lower values calculated within Monte Carlo methods. Comparison to models shows that interaction-induced departure from 2D kinematics is either small or absent. In particular, comparison with a lattice Monte Carlo suggests that kinematics is 2D in the strongly-interacting regime.
\end{abstract}

\maketitle

\section{Introduction}

Thermodynamics of 2D systems is challenging due to bigger than in 3D role of fluctuations. According to the Ginzburg-Levanyuk criterion~\cite{GinzburgLevanyuk1959eng,GinzburgLevanyuk1960eng}, the reduction of the spatial dimensionality limits the applicability of mean-field models, which neglect fluctuations. This may be seen on the example of a system with simple composition, a gas of fermions with \textit{s}-wave attraction.
Description of its ground state using mean field of Cooper pairs~\cite{Miyake1983,Randeria2DCrossover1989prl,MKaganBook2013}, on one hand, predicts qualitatively reasonable behavior of the pairing gap and formation of bosonic molecular dimers as the interaction increases.
On the other hand, the mean-field model gives non-physical answer for the pressure: the Fermi pressure is predicted to be present in the bosonic regime~\cite{Giorgini2D2011}.
Introduction of the order parameter fluctuations into the mean-field model~\cite{SalasnichCompositeBosonsFromFluctuations2015,Fermi2DMeanFieldPlusFluctuations2015} makes the behavior of the pressure qualitatively correct.

Properties of a gas of point-like fermions with \textit{s}-wave attraction may be calculated from first principles.
This, for example, is one of few systems for which the Landau Fermi-liquid theory has been computed from microscopic parameters~\cite{Bloom1975,Miyake1983,FermiLiquid2D1992}.
%
%
Other approaches to calculating the 2D Fermi gas equation of state from first principles, as well as calculating the pressure, include diffusion quantum Monte Carlo~\cite{Giorgini2D2011,DiffusionMC2015}, self-consistent $T$-matrix~\cite{Fermi2DEOSandPressureParish2014}, finite-temperature lattice Monte Carlo~\cite{Fermi2DAbInitioLattice2015}, and auxiliary-field quantum Monte Carlo~\cite{Fermi2DExactGS2015}.

Ultracold gas of Fermi atoms compressed along one direction~\cite{Fermi2D} most accurately corresponds, among other laboratory systems, to the model of the 2D Fermi gas with \textit{s}-wave interactions. An experiment with the ultracold gas is a way of testing the noted theories. Recent experiments have provided data for reconstruction of the equation of state at zero~\cite{FermiBose2DCrossover,Jochim2DThermodynamics2015} and finite temperature~\cite{Vale2DThermodynamics2015,Jochim2DThermodynamics2015}.

In this article, we report on measuring the spatial profiles of the pressure in nonuniform deeply degenerate Fermi gases with \textit{s}-wave attraction and primarily 2D kinematics. For this measurement we developed a method of recovering the 2D pressure profile from a 1D integrated density distribution. The measurement does not require assumptions about the form of the equation of state because the pressure is found by equating the gradient of the pressure to the gradient of the trapping potential~\cite{ThomasUniversal,HoLocalMeas2009}. This paper complements our recent work~\cite{FermiBose2DCrossover}, where the data relevant to the nearly uniform trap center has been reported.
Pressure measurement at different points of a nonuniform cloud gives additional information for comparison with theoretical models. Because the whole spatial pressure is measured within a single experimental shot, the relative pressure at different points has only weak sensitivity to fluctuations of the total atom number and trapping potential.
Since the publication of Ref.~\cite{FermiBose2DCrossover}, several relevant models appeared~\cite{Fermi2DMeanFieldPlusFluctuations2015,Fermi2DEOSandPressureParish2014,Fermi2DAbInitioLattice2015,Fermi2DExactGS2015,DiffusionMC2015}. We compare the pressure profiles to the models that give most distinct results in the regimes of strong and weak interactions. Because of nonuniformity and finite size of the clouds, mesoscopic effects may be present.
We also discuss the effect of the interactions on the kinematic dimensionality.

The paper is organized as follows.
The experimental system is described in Sec.~\ref{sec:ExpSystem}.
Two-body interactions together with their parametrization are discussed in Sec.~\ref{sec:Interaction}.
The method for the measuring the pressure profile is presented in Sec.~\ref{sec:PressureMeasurement}.
Measurements are compared to the models in Sec.~\ref{sec:Comparison}.
The effect of interactions on the kinematic dimensionality is discussed in Sec.~\ref{sec:InteractionEffectOn2D}.
The conclusion is in Sec.~\ref{sec:Conclusion}.

\section{Experimental system}\label{sec:ExpSystem}

\begin{figure}[htb!]
\begin{center}
\includegraphics[width=0.33\linewidth]{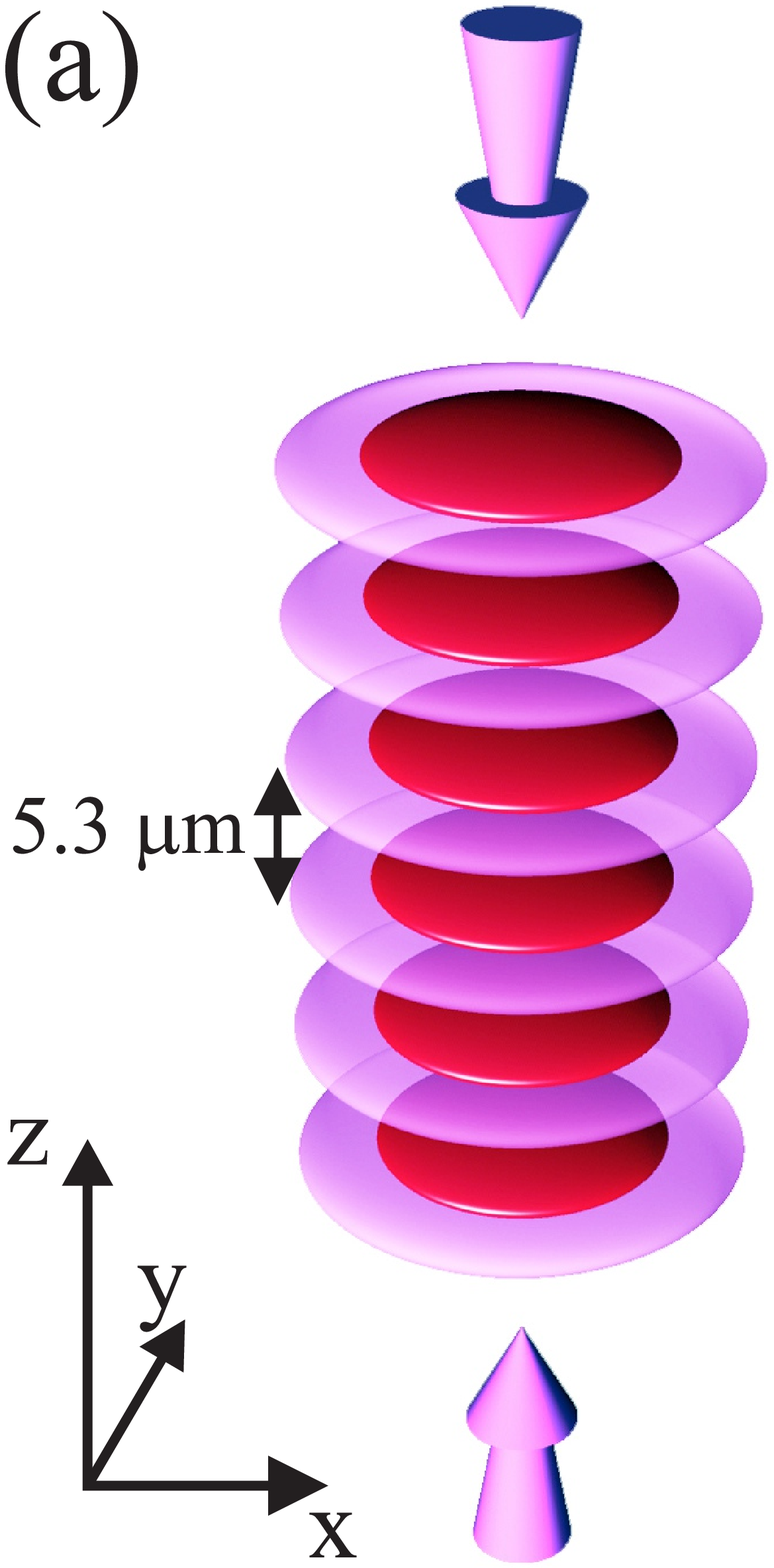}
\includegraphics[width=0.62\linewidth]{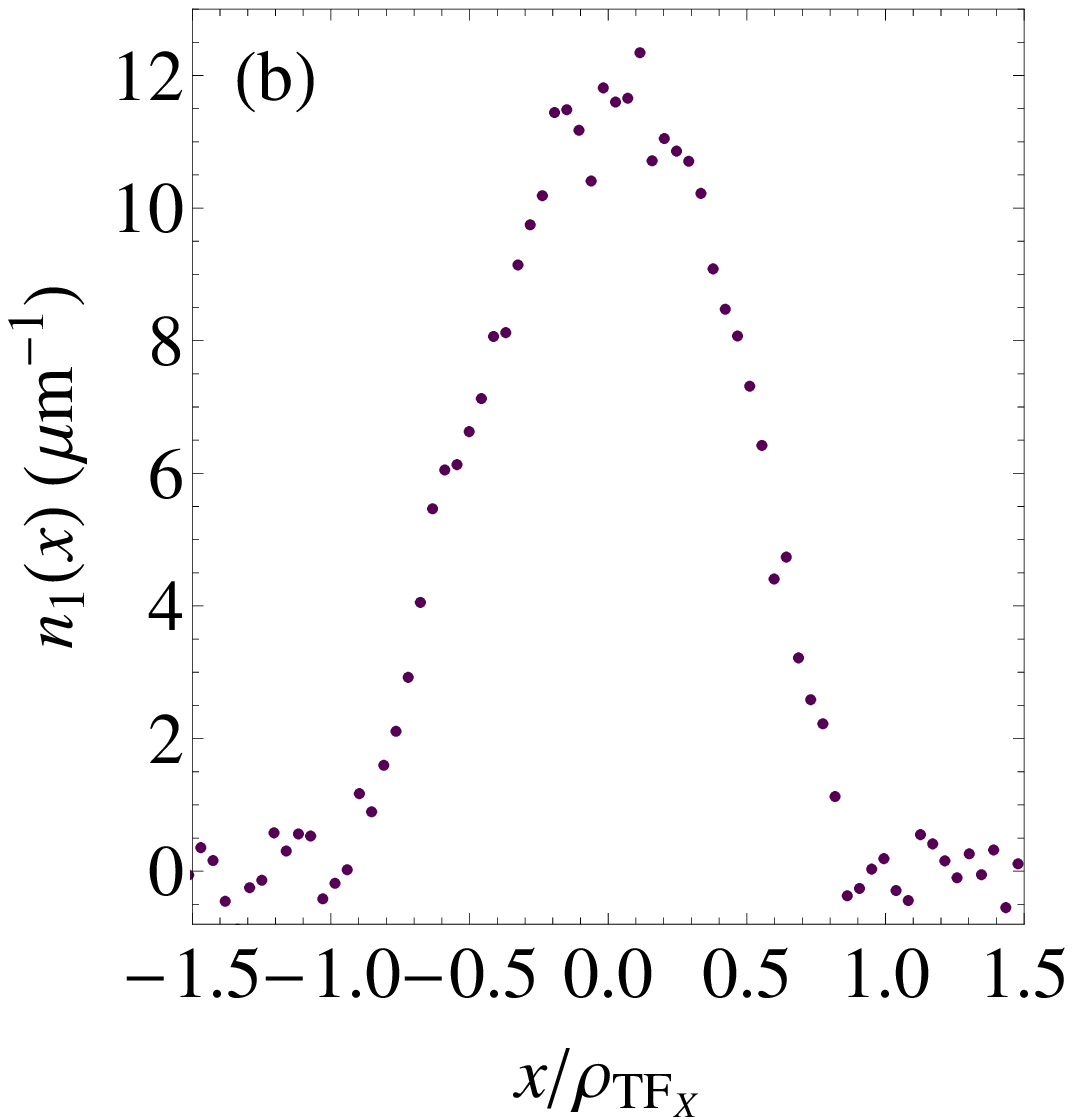}
\end{center}
\caption{(a) Trapping a series of clouds, each being a 2D Fermi gas, in anti-nodes of a standing electromagnetic wave. The gas is shown in dark red, while the intensity of the trapping radiation is shown in light purple. (b) Linear density profile $n_1(x)$ at $B=1000$ G; $\rho_{\text{TF}x}=38.7$ $\mu$m.} \label{fig:TrapAndDensityProfile}
\end{figure}

The experimental setup has been described in~\cite{FermiBose2DCrossover} and references therein. Fermionic lithium-6 atoms equally populate two lowest-energy internal states, $|1\rangle$ and $|2\rangle$, corresponding at zero magnetic field to the states $|F=\frac12,F_z=\frac12\rangle$ and $|F=\frac12,F_z=-\frac12\rangle$. These two states are analogous to the spin-up and spin-down states of electrons in solid materials. A chain of clouds, each being a Fermi gas with nearly 2D kinematics, is created by means of an optical lattice, as schematically shown in Fig.~\ref{fig:TrapAndDensityProfile}(a).
The potential energy of an atom in the lattice is
\begin{equation}
V(x,y,z)=V_0\left[1-\exp\!\left(-\frac{x^2}{\rho_x^2}-\frac{y^2}{\rho_y^2}\right)\,\cos^2kz\right],
\label{eq:OptLattice}
\end{equation}
where $\rho_x$ and $\rho_y$ is the size of the potential along $x$ and $y$ respectively, $k=2\pi/(10.6\text{ $\mu$m})$ is the wave vector of the radiation producing the lattice, and $V_0$ is the potential depth. Near the minima, the potential is close to the harmonic one with frequencies $\omega_x=\sqrt{2V_0/m\rho_x^2}$, $\omega_y=\sqrt{2V_0/m\rho_y^2}$, and $\omega_z=k\sqrt{2V_0/m}$, where $m$ is the atomic mass. Typical values of the frequencies measured by parametric energy input into the ideal gas~\cite{Parametric2015} are $\omega_x/2\pi=92.3\pm0.5$~Hz, $\omega_y/2\pi=139.6\pm0.7$~Hz, and $\omega_z/2\pi=6110\pm120$~Hz. Joint drift of the frequencies between different experiments is less than 1\% and is being accounted for.
The number of atoms $N$ in each cloud varies in range 410--610 per spin state.
Due to the strong anisotropy of the traps, $\omega_z\gg\omega_\perp\equiv\sqrt{\omega_x\omega_y}$, in an ideal Fermi gas with such atom number, all particles can be placed into the lowest state of motion along $z$, which makes kinematics two-dimensional.
%
%
The Fermi energy of 2D noninteracting gas, $E_F$, is found from the equation
\begin{equation}
\hbar^2\omega_x\omega_y N=E_FV_0+(V_0^2-E_FV_0)\ln\left(1-\frac{E_F}{V_0}\right),
\end{equation}
which yields $E_F=(0.53$--$0.65)\hbar\omega_z$. Because of deep degeneracy, thermal excitations do not break 2D kinematics. The effect of interactions on kinematic dimensionality is discussed in Sec.~\ref{sec:InteractionEffectOn2D}.

The degree of degeneracy and, in case of weak interactions, the temperature may be judged from the linear density profile $n_1(x)=\int n_2(x,y)dy$, where $n_2(x,y)$ is the planar density distribution per spin state in a single cloud. An example of measured $n_1(x)$ is shown in Fig.~\ref{fig:TrapAndDensityProfile}(b), where the horizontal scale is normalized to the Thomas-Fermi radius of the ideal 2D Fermi gas, $\rho_{\text{TF}x}=\rho_x\sqrt{-\ln(1-E_F/V_0)}$.
The distribution $n_1(x)$ is obtained by imaging the clouds along the $y$ direction, which effectively integrates the planar density along $y$, and averaging over 30 adjacent clouds, thereby reducing noise and fluctuations. Fitting $n_1(x)$ by the Thomas-Fermi profile of the ideal Fermi gas~\cite{FermiBose2DCrossover} gives dimensionless temperature parameter $(T/E_F)_{\text{fit}}$ in range 0.00--0.12 indicating deep degeneracy. For weakly interacting Fermi gases this parameter coincides with $T/E_F$, temperature in $E_F$ units.

\section{Two-body interactions}\label{sec:Interaction}

Due to low temperature and large interparticle distances only \textit{s}-wave interactions are possible, whose value is tuned using a Fano-Feshbach resonance~\cite{FeshbachReview2010}, by placing the gas into an external nearly uniform magnetic field $B$. The field $B$ is chosen in range 1400--810~G including the resonance at 832~G \cite{JochimNewLiFeshbach2013}. In a 3D gas these magnetic fields would correspond to the inverse 3D scattering length $1/a$ in range from $-1/2790$ to $1/16960$ Bohr$^{-1}$.

Theoretical models largely use the 2D \textit{s}-wave scattering length $a_2$, which appears in the problem of scattering on a purely 2D potential that is $z$-independent. In the experiment, the scattering is quasi-2D (Q2D), i.~e., kinematically 2D particles interact on 3D potentials of a negligibly small radius. The 2D and Q2D problem may be mapped onto each other because the 2-body wavefunction at large distances takes same form in both cases:
\begin{equation}
\psi_\perp(\vec{\rho'})\simeq e^{i\vec q\cdot\vec{\rho'}}-f\frac{e^{iq\rho'-i\pi/4}}{\sqrt{8\pi q\rho'}},
\end{equation}
where $\vec{\rho'}$ is the planar vector between the two atoms and $\hbar\vec q\equiv(\vec p_1-\vec p_2)/2$ is the scattering momentum expressed via the atomic momenta in the laboratory reference frame, $\vec p_1$ and $\vec p_2$.
Only the amplitude $f$ depends on whether the scattering is 2D or Q2D. For the scattering on a hypothetical 2D potential,
\begin{equation}
f=f_{\text{2D}}(q,a_2)\equiv-\frac{2\pi}{\ln(qa_2e^\gamma/2i)},
\label{eq:f2}
\end{equation}
where $\gamma\simeq0.577$ is Euler's constant. For the actual Q2D scattering~\cite{Shlyapnikov2DScattering2001},
%
%
\begin{equation}
f=f_{\text{Q2D}}(q,a,l_z)\equiv\frac{2\pi}{\sqrt\pi l_z/a+w(q^2l_z^2)/2},
\label{eq:f2ho}
\end{equation}
where $l_z=\sqrt{\hbar/2m\omega_z}$ is the size of the state along the quantized direction of the Q2D problem and function $w(\xi)$ is defined by the limit
\begin{equation}
w(\xi)\!\equiv\!\lim_{J\rightarrow\infty}\!\left[\sqrt{\frac{4J}\pi}\ln\frac J{e^2}- \sum_{j=0}^J\frac{(2j-1)!!}{(2j)!!}\ln(j-\xi-i0)\right]\!\!.
\label{eq:wOfxi}
\end{equation}
As a result the value of $a_2$ of the corresponding 2D problem is found from the equation~\cite{FermiBose2DCrossover,UFN2016eng}:
\begin{equation}
f_{\text{Q2D}}(q,a,l_z)=f_{\text{2D}}(q,a_2).\label{eq:Finda2}
\end{equation}
In the limit $ql_z\ll1$, Eq.~(\ref{eq:Finda2}) gives known result $a_2\simeq2.96\,l_z\,e^{-\sqrt\pi l_z/a}$ \cite{Shlyapnikov2DScattering2001,BlochLowDReview2008}. It shows that by changing $a$ via the Feshbach resonance and keeping $l_z$ fixed, one may tune the 2D scattering length.
In a many-body problem, the momentum $\hbar q$ differs from 0 and may be estimated as $\hbar q=\sqrt{\mathstrut2\mu m}$ from the chemical potential $\mu$ that does not include the two-body binding energy~\cite{FermiBose2DCrossover}. This estimate is exact for deeply degenerate weakly interacting Fermi gases where the colliding particles are on the Fermi surface. After the 2D and Q2D scattering problems are related to each other, one may compare measurement to predictions of purely 2D models.

\section{Pressure profile measurement}\label{sec:PressureMeasurement}

The measurement of the pressure distribution in the clouds is based on the force balance equation
\begin{equation}
\nabla_\perp P(x,y)=-n_2(x,y)\nabla_\perp V(x,y,0),\label{eq:PressureBalance}
\end{equation}
where $P$ is the partial pressure of each spin component. The planar density profile $n_2(x,y)$ may in principle be recovered from the integral $n_1(x)$ due to the cylindrical symmetry of potential~(\ref{eq:OptLattice}) in stretched coordinates $(x,\tilde{y}\equiv y\,\omega_y/\omega_x)$. The inverse Abel transform yields
\begin{equation}
n_2(\tilde{\rho})=-\frac{\omega_y/\omega_x}\pi\int_{\tilde{\rho}}^\infty\frac{dn_1(x)}{dx}\frac{dx}{\sqrt{x^2-\tilde{\rho}^2}},
\label{eq:InvAbel}
\end{equation}
where $\tilde\rho\equiv\sqrt{x^2+\tilde{y}^2}$. The actual calculation of $n_2(\tilde\rho)$ should be avoided, however, because the derivative $dn_1/dx$ enters the transform together with the divergent denominator, which all together strongly amplify high-spatial-frequency noise. This noise primarily comes from imaging: The first and second biggest contributors are the photon shot noise and the charge-coupled-device readout noise respectively.

The pressure at any point of the cloud may be obtained from $n_1(x)$ directly by substituting (\ref{eq:InvAbel}) into (\ref{eq:PressureBalance}) and integrating, which yields:
\begin{align}\label{eq:P2vian1}
P(\tilde{\rho})=&
\frac{m\omega_\perp^2e^{-{\tilde\rho}^2/\rho_x^2}}\pi  \left[\int_{\tilde\rho}^\infty\frac{n_1(x)\,x\,dx}{\sqrt{x^2-{\tilde\rho}^2}}-\right.\nonumber\\
&\left.\frac2{\rho_x}\int_{\tilde\rho}^{\infty}n_1(x)\,x\, D_+\!\!\left(\frac{\sqrt{x^2-{\tilde\rho}^2}}{\rho_x}\right)dx\right],
\end{align}
where $D_+$ designates the Dawson function:
\begin{equation}
D_+(\xi)\equiv e^{-\xi^2}\int_0^\xi e^{\eta^2}d\eta.
\end{equation}
Neither $n_2(\tilde\rho)$ nor $dn_1/dx$ enter formula~(\ref{eq:P2vian1}). As a result, the noise is kept at a reasonable level as it will be seen in the further data analysis. Formula (\ref{eq:P2vian1}) as well as the pressure measurement method is applicable without any limits on temperature or composition of the trapped gas.

\begin{figure}[htb!]
\begin{center}\includegraphics[width=1.0\linewidth]{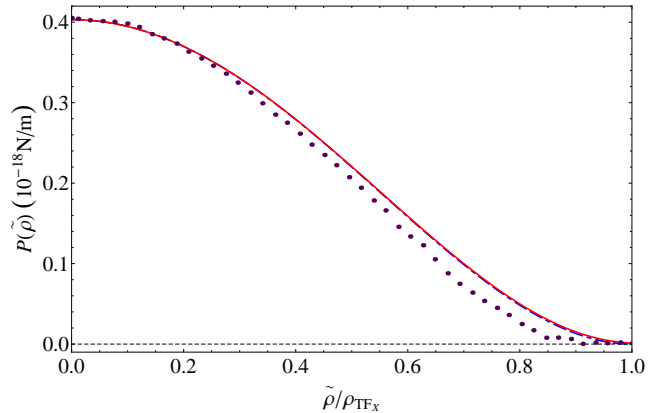}\end{center}
\caption{Pressure profile in a trapped gas. Dots are the data corresponding to the density distribution in Fig.~\ref{fig:TrapAndDensityProfile}(b). Blue dash-dash-dotted curve and red solid curve are the theory of the ideal gas respectively at $T=0$ and $T=0.04E_F$, where the latter is the temperature in the experiment.}\label{fig:PressureProfile}
\end{figure}

An example of measured pressure profile is shown in Fig.~\ref{fig:PressureProfile}.
The theory of the ideal Fermi gas in the Thomas-Fermi approximation is plotted for comparison. The interactions leave the central pressure nearly unchanged because the potential is close to the harmonic one, where the pressure at the origin is $m\omega_\perp^2N/2\pi$ regardless of the interactions and temperature. Meanwhile the interactions show up away from the center where the difference with the ideal-gas profile is evident.

Since the difference with the ideal-gas pressure profile is not large, detection of the interaction effects requires precise pressure measurement. The precision may be tested via two boundary conditions:
(i) $P(\tilde\rho)$ should drop to 0 as $\tilde\rho\rightarrow\infty$ and (ii) at the origin $\tilde\rho=0$, the pressure should take value
\begin{equation}
P(0)=\frac{m\omega_\perp^2N}{2\pi}\left(1-\frac{m\omega_x^2\left\langle x^2\right\rangle}{V_0}\right),
\end{equation}
where $\left\langle x^2\right\rangle\equiv\frac1N\int x^2n_1(x)dx$ and the last term is the small anharmonic correction. The noise in $n_1(x)$ does not allow to meet these two conditions exactly.
In order to satisfy either of these conditions, we apply a small constant shift to $P(\tilde\rho)$. Satisfying each condition separately gives slightly different shifts. For each experimental run, the mean of these two shifts is included in $P(\tilde\rho)$. Such shift is $(0.1\pm0.8)\%$ of $P(0)$ when averaged over all experiments.
The half-difference between these two shifts equals $0.4\%$ of $P(0)$ on average and is included into the systematic errors.

Profiles $n_1(x)$ used in this paper are a sampling of profiles, which have been analyzed in Ref.~\cite{FermiBose2DCrossover} with the purpose of finding density and pressure in the almost uniform central part of the cloud. This sampling is chosen on the condition of nearly identical parameters of the potential, smallness of the temperature parameter $(T/E_F)_{\text{fit}}$, and closeness of the number of atoms $N$. Only one parameter is significantly varied, the magnetic field, which is responsible for the value of the interparticle interaction. Thereby, the number of parameters for comparing with theoretical models has been minimized. Favorable signal-to-noise ratio allows for comparison to a number of theoretical models.

\section{Comparison of pressure profiles to theoretical models}\label{sec:Comparison}

\begin{figure*}[htb!]
\begin{center}
\includegraphics[height=39mm]{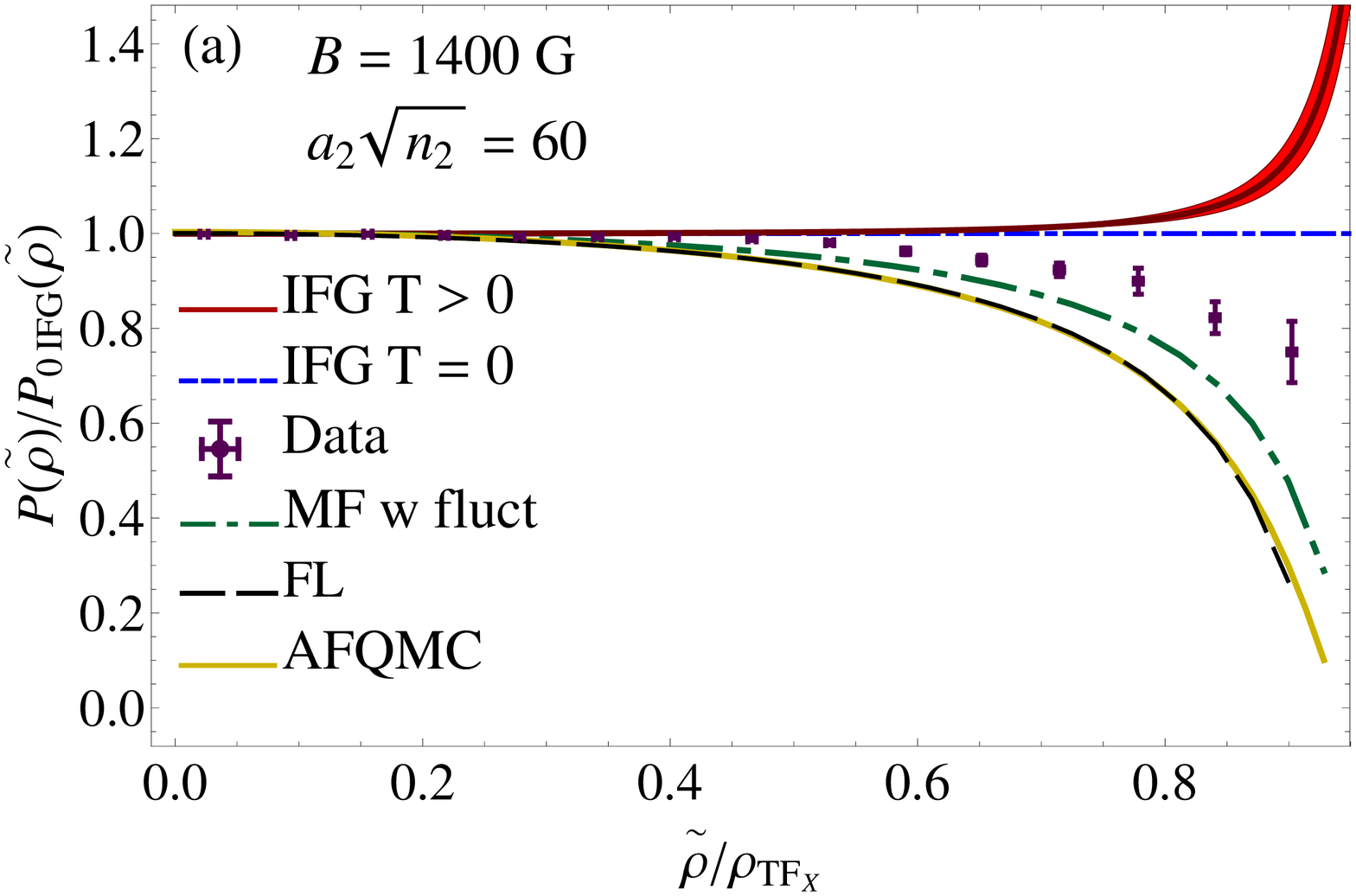}
\includegraphics[height=39mm]{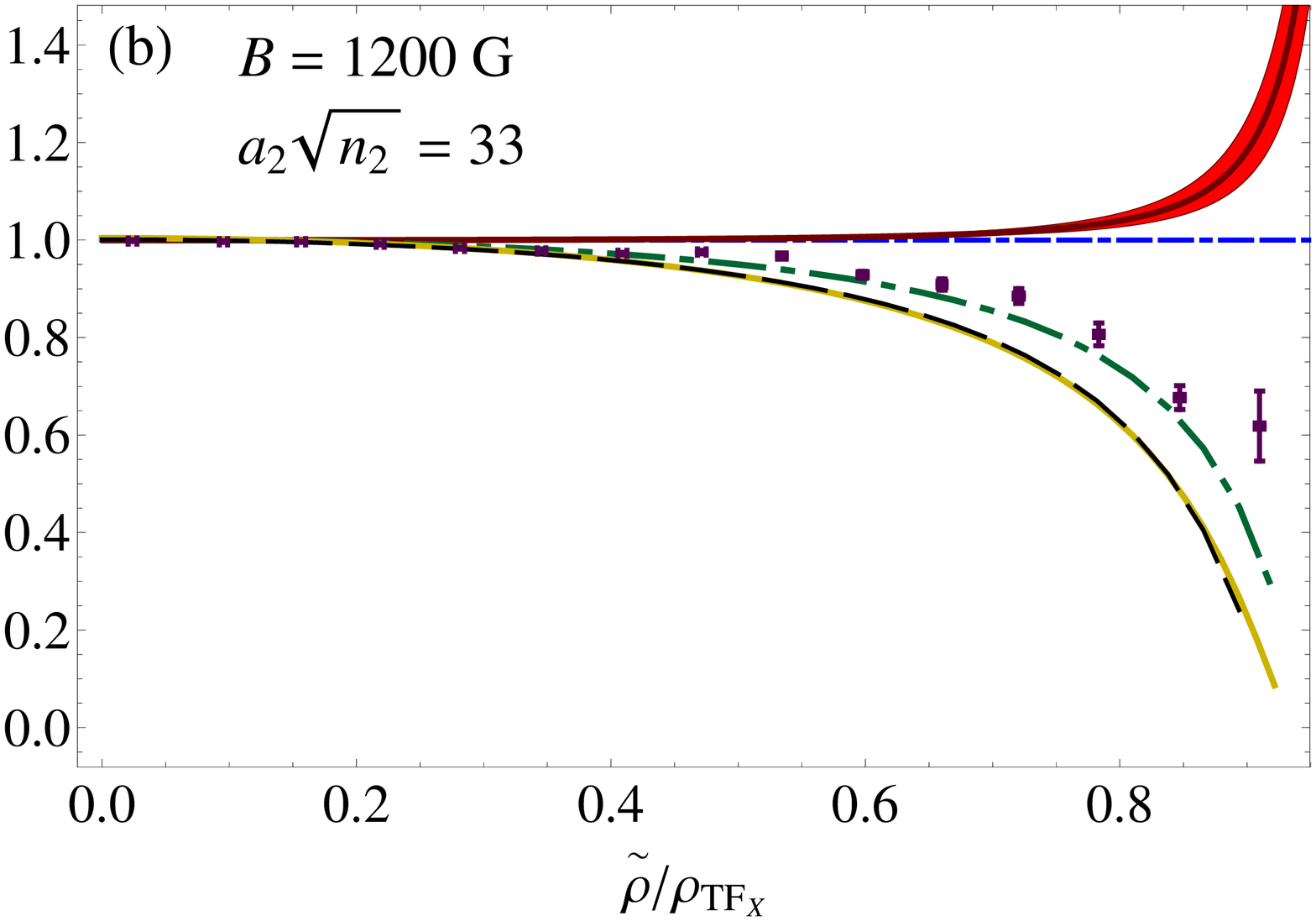}
\includegraphics[height=39mm]{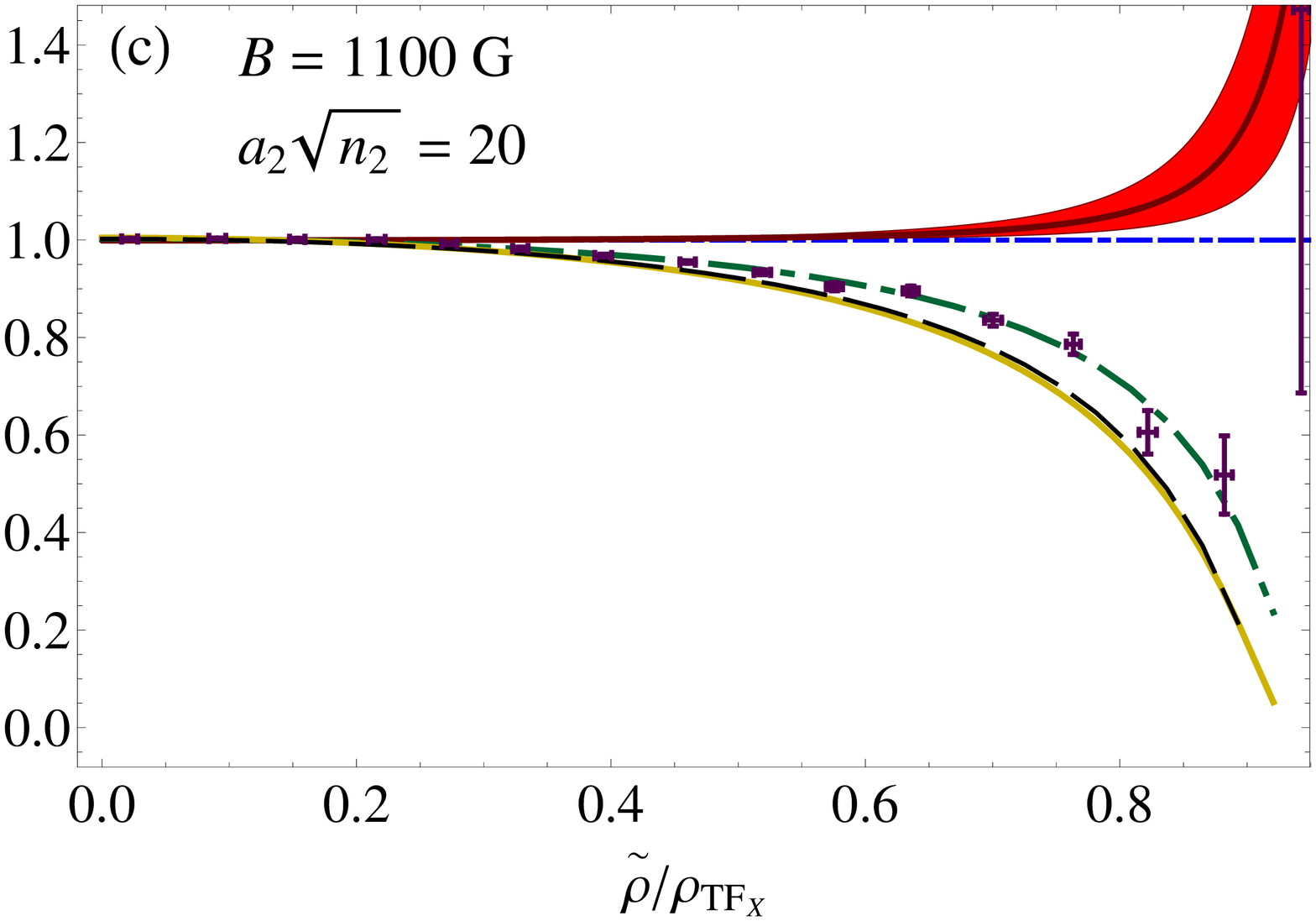}\vspace{1mm}
\includegraphics[height=39mm]{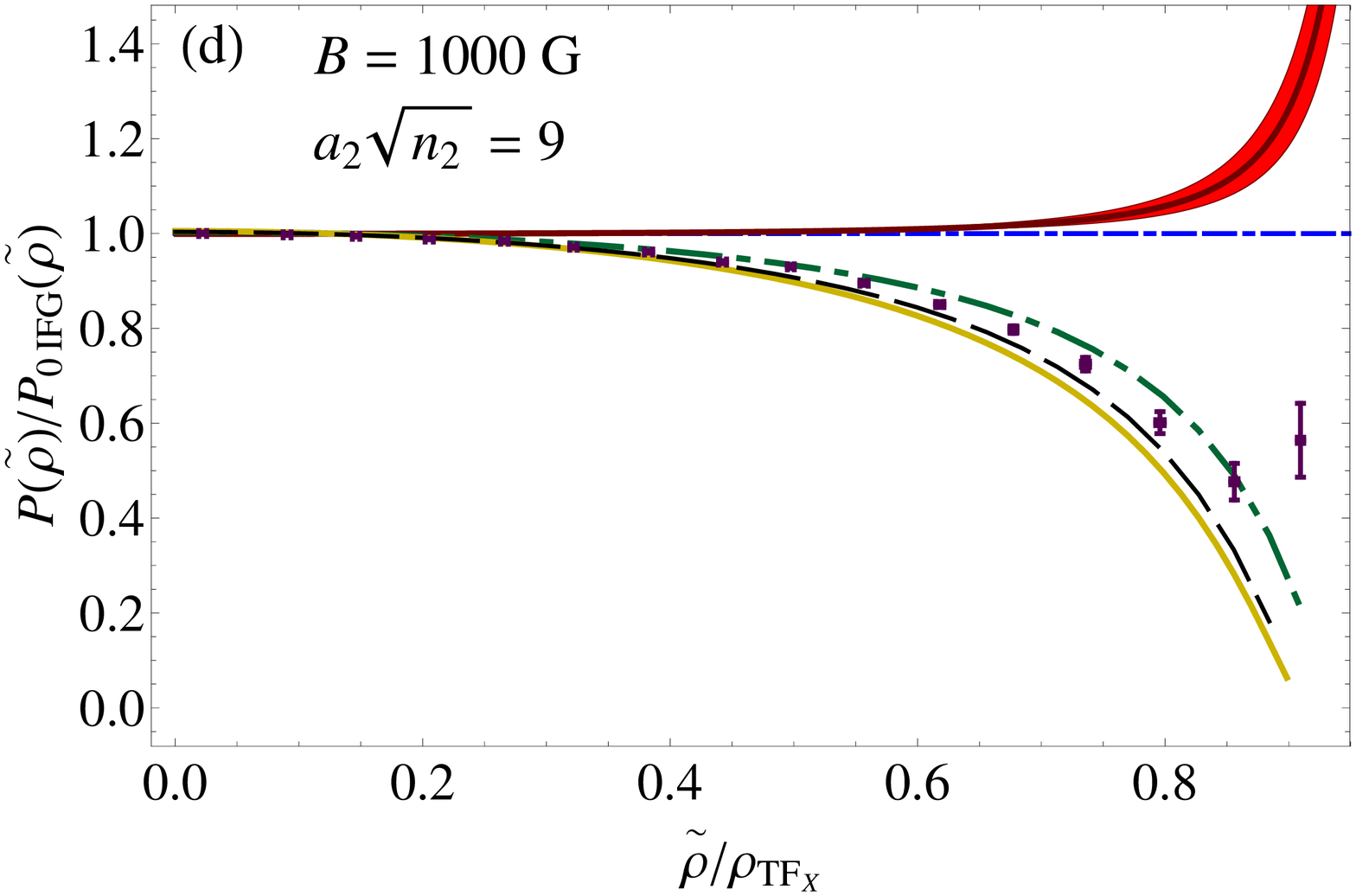}
\includegraphics[height=39mm]{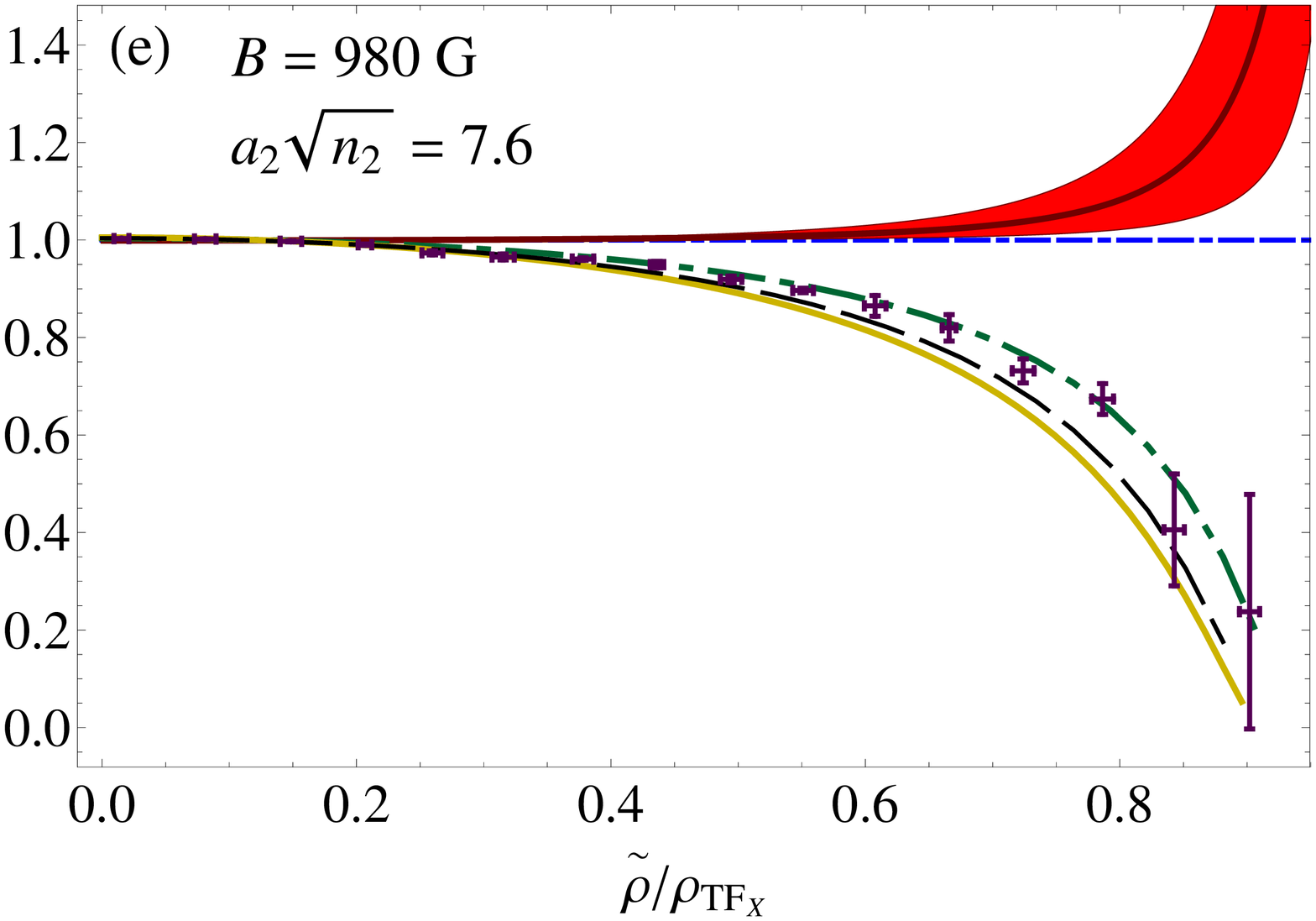}
\includegraphics[height=39mm]{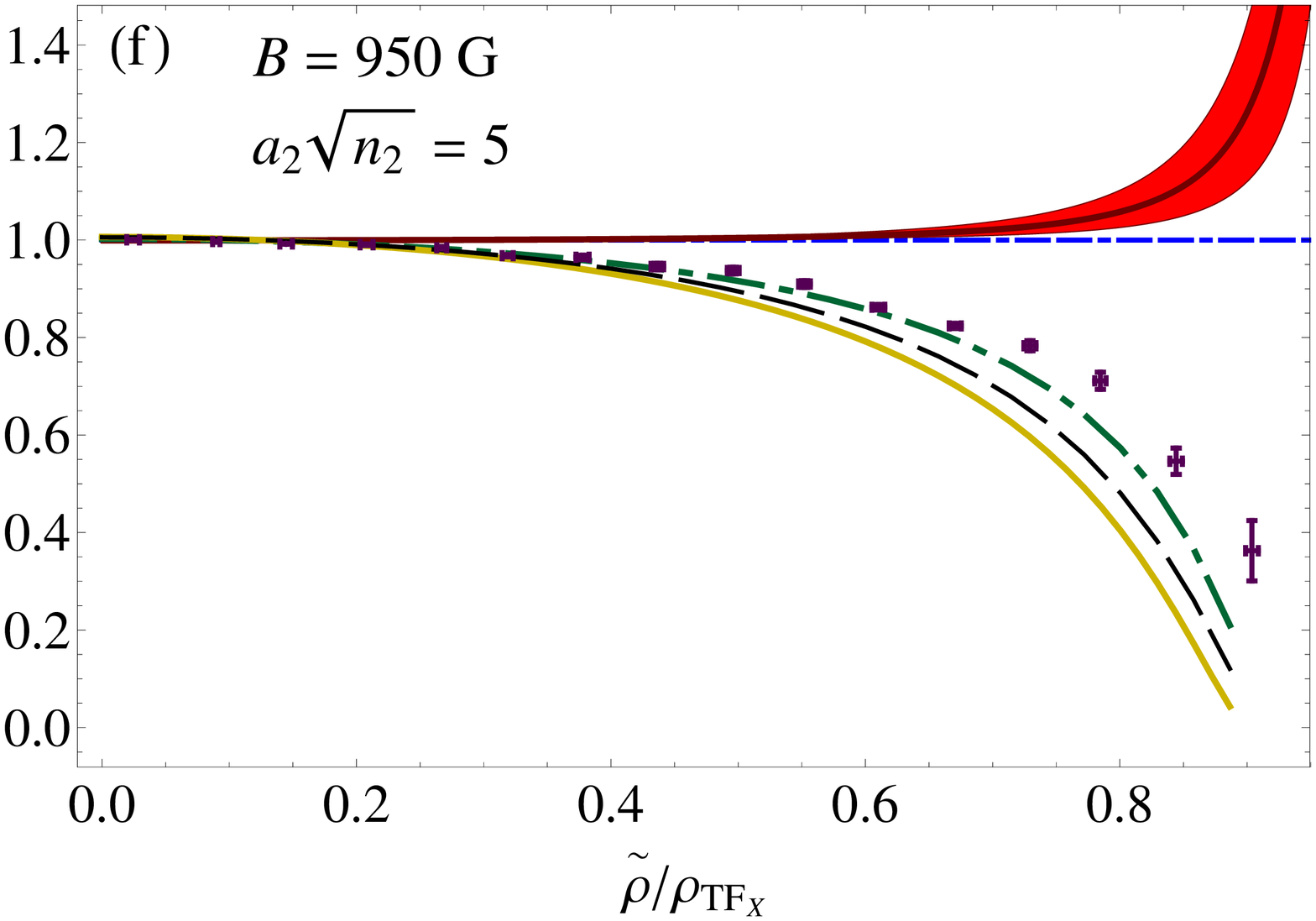}\vspace{1mm}
\includegraphics[height=39mm]{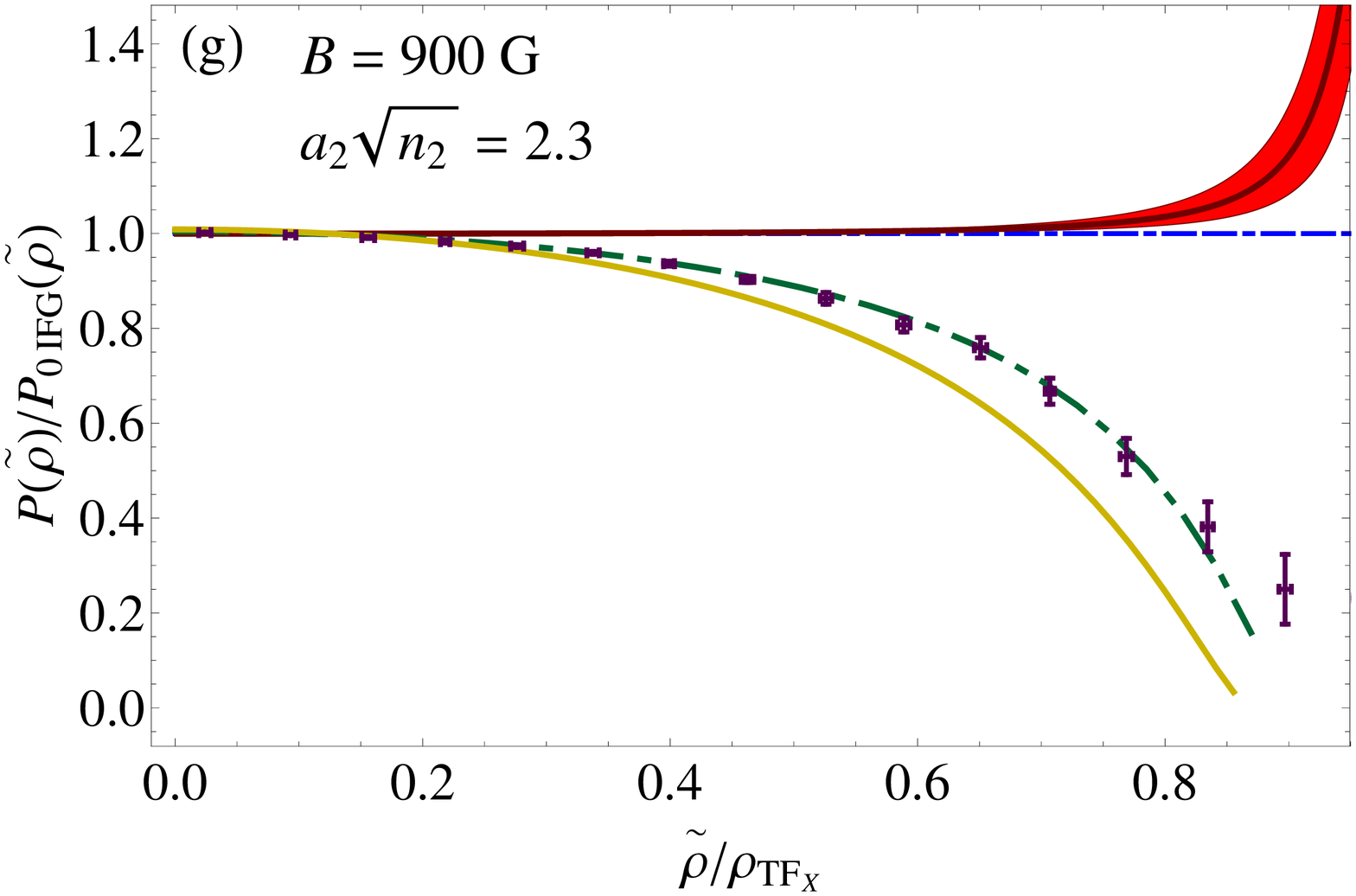}
\includegraphics[height=39mm]{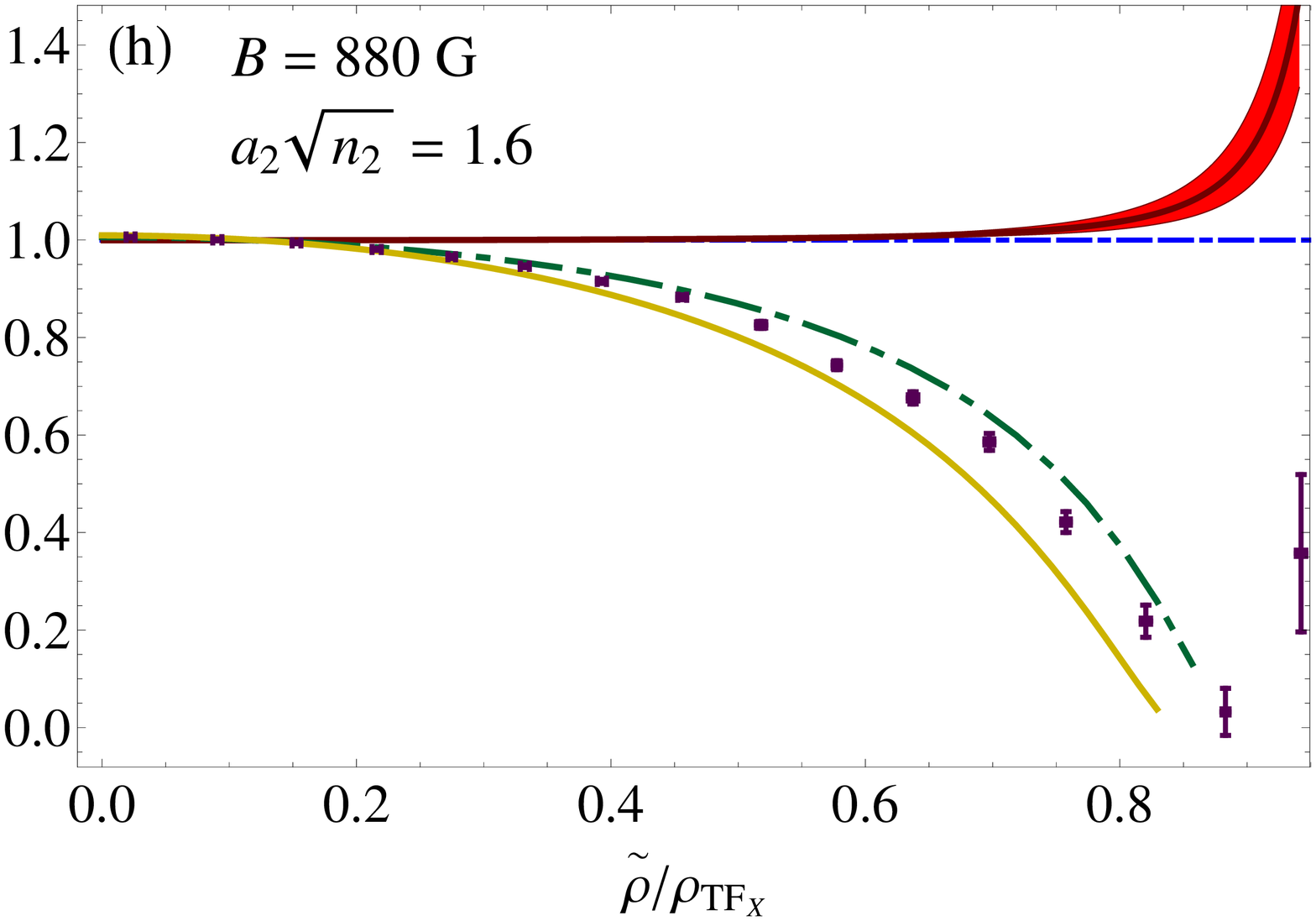}
\includegraphics[height=39mm]{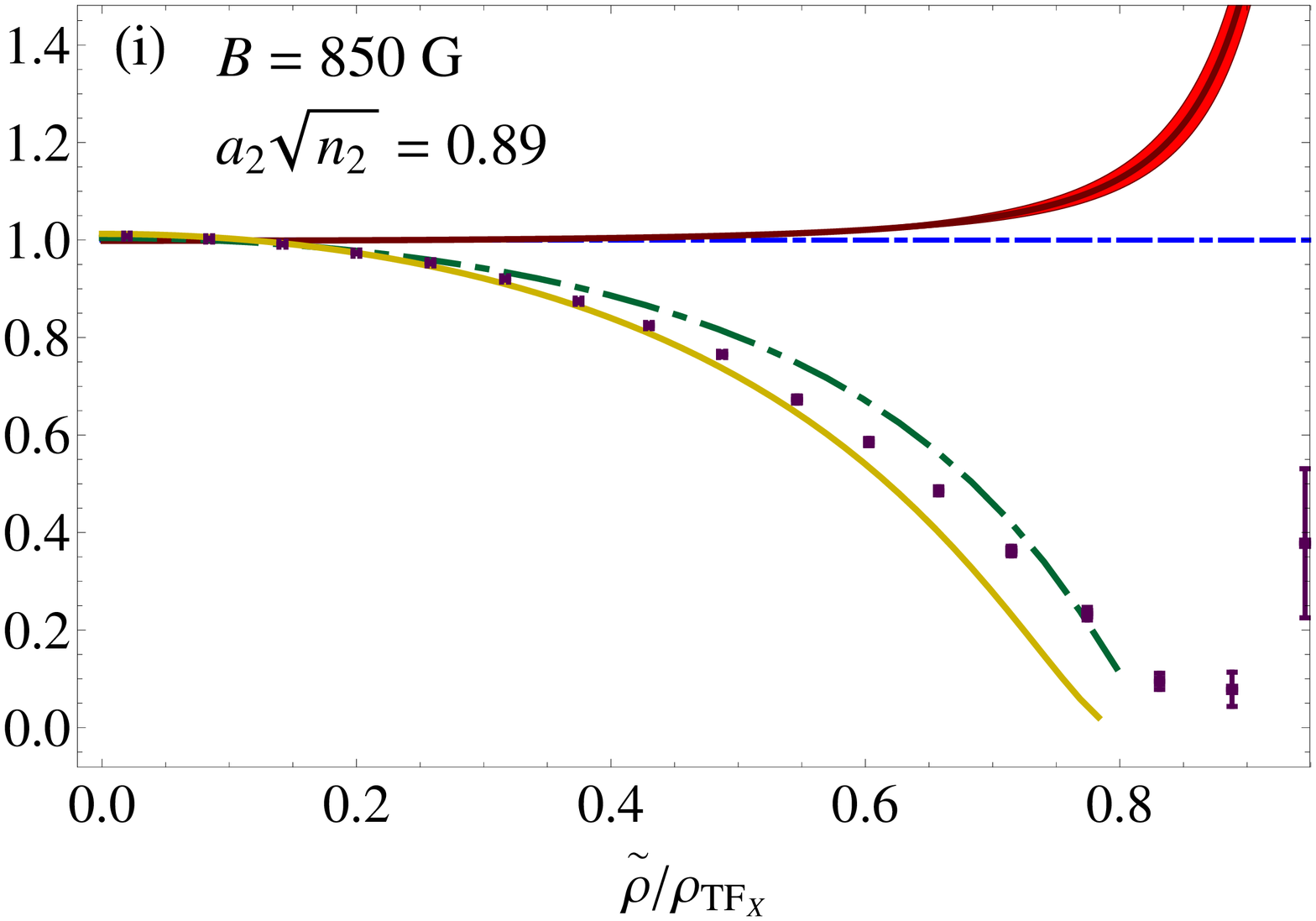}\vspace{1mm}
\includegraphics[height=39mm]{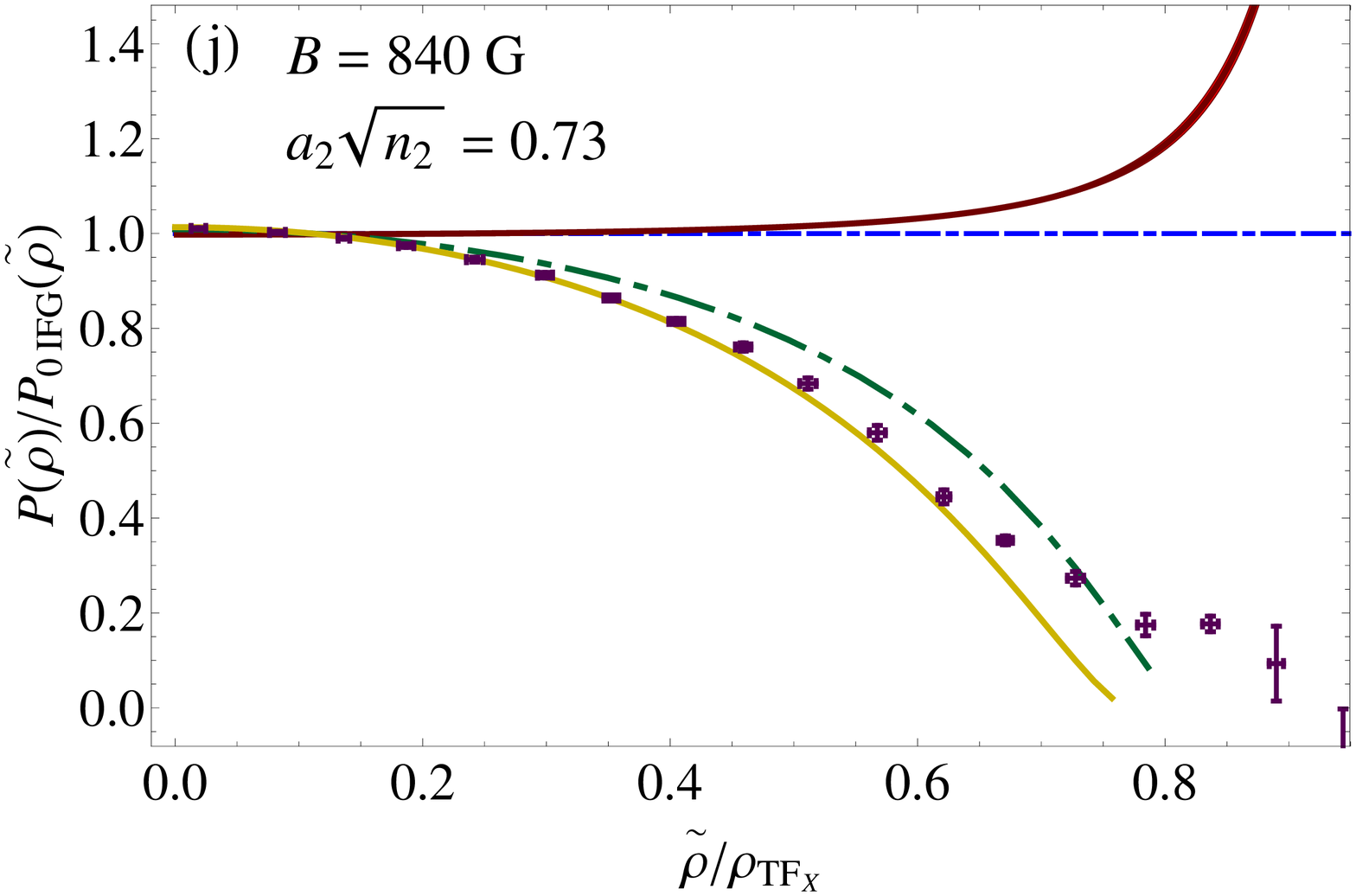}
\includegraphics[height=39mm]{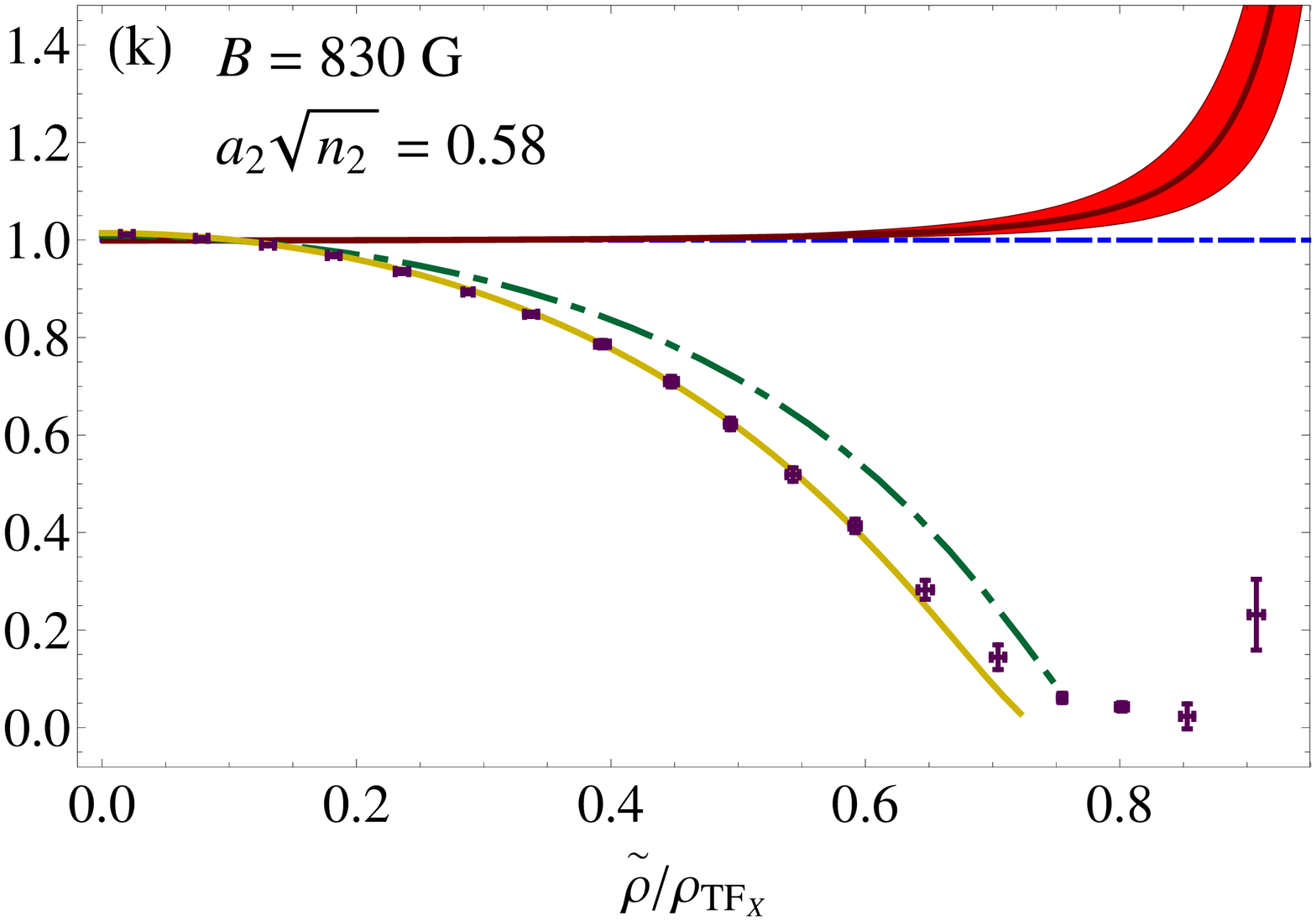}
\includegraphics[height=39mm]{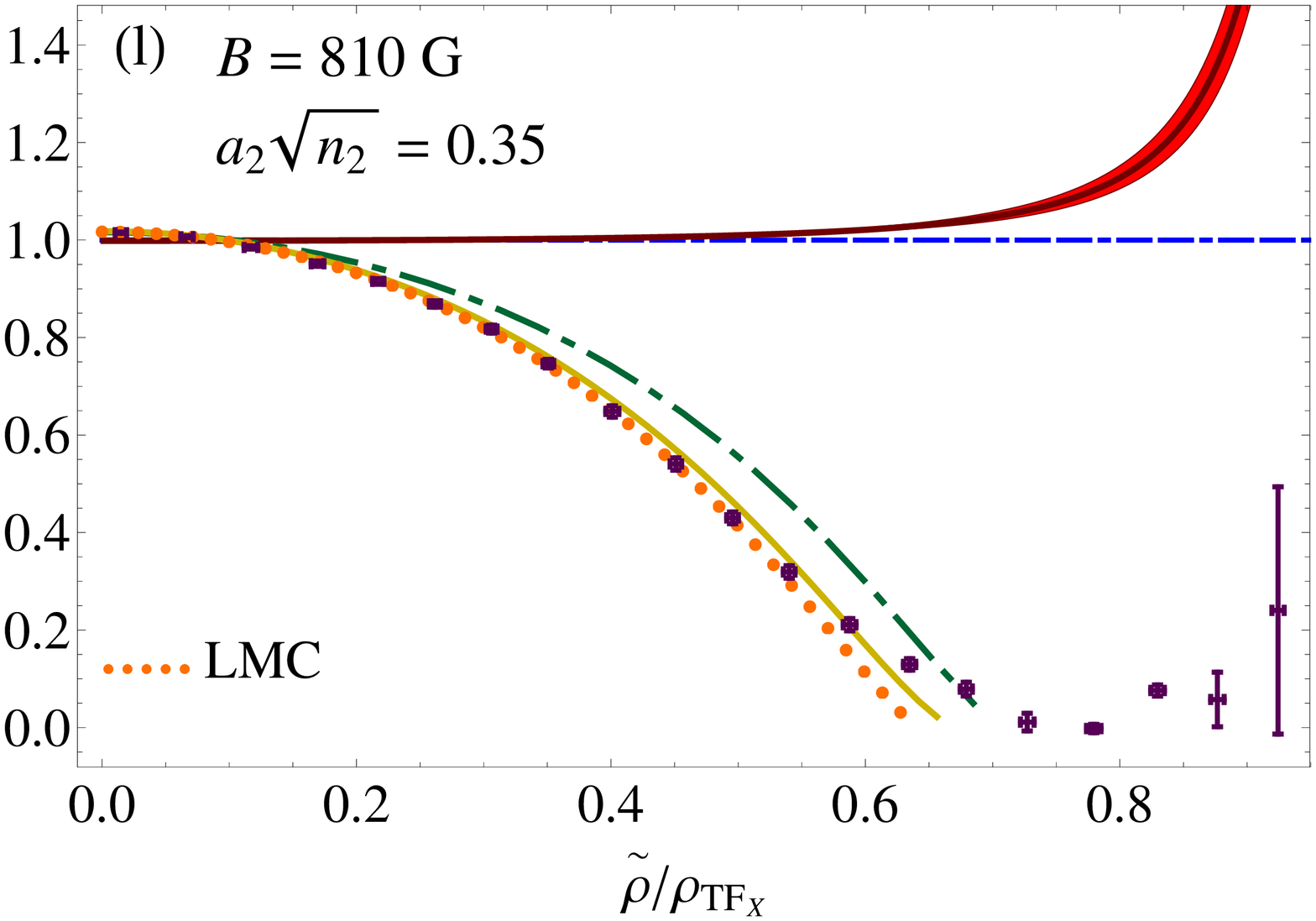}
\end{center}
\caption{Pressure profiles $P(\tilde\rho)$ normalized to $P_{0\,\text{IFG}}(\tilde\rho)$, the ideal Fermi gas profile at $T=0$. In each panel, dimensionless interaction parameter $a_2\sqrt{n_2}$ at the cloud center and the magnetic field are marked. Dots are the data, horizontal blue dash-dash-dotted line is the ideal Fermi gas (IFG) model at $T=0$ as well as the mean-field model~\cite{Randeria2DCrossover1989prl}.
Green dash-dotted curve is the mean-field (MF) model with fluctuations~\cite{Fermi2DMeanFieldPlusFluctuations2015},
dashed black curve is the Fermi-liquid (FL) theory~\cite{Bloom1975},
yellow solid curve is the auxiliary-field quantum Monte Carlo (AFQMC) \cite{Fermi2DExactGS2015}.
Red shading with burgundy curve inside is the finite-temperature IFG theory, where the curve is drawn for the mean $(T/E_F)_{\text{fit}}$ observed in the experiments at the respective $B$, while the shading reflects the uncertainty of $(T/E_F)_{\text{fit}}$.
In panel (l), finite-temperature lattice Monte Carlo (LMC) \cite{Fermi2DAbInitioLattice2015} at $T=0$ is shown by the orange dotted curve; for panels (e--k) the LMC agrees with AFQMC within line thickness and not shown.}\label{fig:ComparePressure}
\end{figure*}

Comparison of the data to calculations at different interaction values is done in Fig.~\ref{fig:ComparePressure}.
For emphasizing the differences, both the data and theoretical profiles are normalized to $P_{0\,\text{IFG}}(\tilde\rho)$, the profile of the ideal Fermi gas in the same potential at $T=0$ in the Thomas-Fermi approximation. In each panel, one may see the corresponding magnetic field $B$ and the dimensionless interaction parameter at the cloud center, $a_2\sqrt{n_2}$. In this parameter, the scale of the corresponding purely 2D two-body scattering problem is divided by the mean interparticle distance. Values $a_2\sqrt{n_2}\gg1$ and $a_2\sqrt{n_2}\sim1$ correspond to the weak and strong interactions respectively, where the border may be drawn at $a_2\sqrt{n_2}=5$. An extended set of parameters relevant to each panel of Fig.~\ref{fig:ComparePressure} is shown in Table~\ref{tbl:ComparePressure}.
\begin{table}{
\begin{tabular}[t]{|c|c|c|c|c|}
\hline
\parbox{4em}{Panel of Fig.~\ref{fig:ComparePressure}} & $B$ (Gauss) & \parbox{5em}{$a_2\sqrt{n_2}$ at the center} & $l_z/a$ & \parbox{6em}{$n_2/n_{2\text{ IFG}}$ at the center} \\ \hline
(a) & 1400 & 60   & $-2.8$ & $1.03\pm0.03$ \\ \hline
(b) & 1200 & 33   & $-2.5$ & $1.06\pm0.07$ \\ \hline
(c) & 1100 & 20   & $-2.2$ & $1.09\pm0.01$ \\ \hline
(d) & 1000 & 9   & $-1.60$ & $1.12\pm0.04$ \\ \hline
(e) &  980 & 7.6  & $-1.38$ & $1.16\pm0.01$ \\ \hline
(f) &  950 & 5.0  & $-1.36$ & $1.13\pm0.04$ \\ \hline
(g) &  900 & 2.3  & $-0.92$ & $1.19\pm0.08$ \\ \hline
(h) &  880 & 1.6  & $-0.69$ & $1.24\pm0.03$ \\ \hline
(i) &  850 & 0.89 & $-0.29$ & $1.39\pm0.08$ \\ \hline
(j) &  840 & 0.73 & $-0.130$ & $1.52\pm0.04$ \\ \hline
(k) &  830 & 0.57 & $0.038$ & $1.63\pm0.08$ \\ \hline
(l) &  810 & 0.35 & $0.41$ & $2.02\pm0.09$ \\ \hline
\end{tabular}}
\caption{Respective parameters for the data of Figs.~\ref{fig:ComparePressure}(a--l). Here $n_{2\text{ IFG}}$ is the 2D numerical density at the center for an ideal Fermi gas with the same atom number and $(T/E_F)_{\text{fit}}$.}\label{tbl:ComparePressure}
\end{table}

Five zero-temperature models are chosen for comparison. Three of them have significantly different predictions in the regime of weak interactions, at $B\geqslant950$~G.
The first one is the mean-field model of Cooper pairs~\cite{Miyake1983,Randeria2DCrossover1989prl,MKaganBook2013}, whose prediction coincides with that of the ideal Fermi gas model $P_{0\,\text{IFG}}(\tilde\rho)$ and is shown by the horizontal line in Figs.~\ref{fig:ComparePressure}.
The second model is the mean-field theory supplemented by order-parameter fluctuations~\cite{Fermi2DMeanFieldPlusFluctuations2015}. One may see that the fluctuations qualitatively change the normalized pressure profile.
The third model is the Fermi-liquid theory~\cite{Bloom1975}, which is constructed for weak interactions only and is inapplicable for $a_2\sqrt{n_2}\sim1$. The Fermi-liquid theory is complemented with the forth model, auxiliary-field quantum Monte Carlo~\cite{Fermi2DExactGS2015}, which is applicable in the whole range of $a_2\sqrt{n_2}$ and is close to the Fermi-liquid theory in the weak-attraction regime. The fifth model, finite-temperature lattice Monte Carlo~\cite{Fermi2DAbInitioLattice2015}, is shown in panel (l) only, because for panels (e--k) its predictions are within line thickness from model~\cite{Fermi2DExactGS2015}, while for $a_2\sqrt{n_2}>8$ [panels (a--d)] numerical results are unavailable.

Within each model, the pressure is calculated in the local density approximation. The calculation is done on the basis of the equation of state, number of particles, and the potential $V(x,y,0)$. Scattering length $a_2$ depends on coordinate $\tilde\rho$ and is being calculated self-consistently with the density profile. The weak dependence of $a_2$ on $\tilde\rho$ comes from the dependence on local chemical potential $\mu$ as seen from Eq.~(\ref{eq:Finda2}) and the subsequent estimate for momentum $\hbar q$.

Data in graphs~\ref{fig:ComparePressure}(a--l) are averaged over 2--15 repetitions of the experiment. Each error bar is the standard error of the mean.

The temperature in each experiment is $\ll E_F$ but different from 0. The estimate for finite temperature effects is shown in Figs.~\ref{fig:ComparePressure}: By the burgundy curves we display the theory of the ideal Fermi gas at the same value of $(T/E_F)_{\text{fit}}$, as on average in experiments at a given $B$, while the red shading corresponds to the standard error of $(T/E_F)_{\text{fit}}$. The excess of the curve over 1 shows the role of finite temperature. One may see that in the major part of the cloud the effect of the temperature is small. Therefore, the difference between the data and the models may be related to the nonzero temperature in a narrow cloud edge only.

The temperature is above the Cooper-pair breaking temperature~\cite{Shlyapnikov2DFermi2003} except for maybe experiments with the strongest attraction. Despite that, the use of models based on Cooper pairs is justified because the pressure does not jump at the phase transition.

For a Fermi gas with weak attraction, the data and models are presented in Figs.~\ref{fig:ComparePressure}(a-f).
In all cases, the data are below the ideal-gas and mean-field theories. Also the data are systematically above the predictions of the Fermi-liquid theory~\cite{Bloom1975,FermiLiquid2D1992}. The mean-field model supplemented by fluctuations~\cite{Fermi2DMeanFieldPlusFluctuations2015} gives the curve, which is the closest to the data.

For the weakest interactions, the finite size of the clouds may affect the interactions.
For the data of Figs.~\ref{fig:ComparePressure}(a,b), the 2D scattering length at the cloud center is $a_2=120$ $\mu$m and $a_2=75$ $\mu$m respectively, i.~e., the scattering length well exceeds the Thomas-Fermi radius $\rho_{\text{TF}x}=39$ $\mu$m. This may potentially inhibit the interaction.
For the data of Fig.~\ref{fig:ComparePressure}(a), such mesoscopic effect could be the reason why the pressure is higher than predicted by the model of Ref.~\cite{Fermi2DMeanFieldPlusFluctuations2015} (green dash-dotted curve).
%
%
The size of the one-body wave function $l_x=\sqrt{\hbar/2m\omega_x}=3$ $\mu$m is notable in comparison with $\rho_{\text{TF}x}$. This one-body quantum effect, however, does not cause any measurable increase of the pressure with respect to calculation in the Thomas-Fermi approximation.

The data of graphs~\ref{fig:ComparePressure}(f--l) taken at $B=950$--$810$~G correspond to strong interactions.
In response to the increasing attraction, the pressure shifts below the model of Ref.~\cite{Fermi2DMeanFieldPlusFluctuations2015}, where the mean field of Cooper pairs is supplemented with fluctuations of the order parameter.
The pressure goes down to the values predicted by the auxiliary-field quantum Monte Carlo~\cite{Fermi2DExactGS2015}. In the case of the largest pair binding, Fig.~\ref{fig:ComparePressure}(l), the data lie slightly below this model and match with the finite-temperature lattice Monte Carlo~\cite{Fermi2DAbInitioLattice2015} drawn for $T=0$.
In the strongly-interacting case, since $a_2\sqrt{n_2}\sim1$, the finite system size should not have any effect on the interactions.

\section{Effect of interactions on kinematic dimensionality}\label{sec:InteractionEffectOn2D}

For comparison with 2D models, it is important that the experimental system remains two-dimensional. In this section, the effect of interactions on kinematic dimensionality is discussed.

Two-body interaction mixes states of the $z$-motion of an atom.
From the standpoint of a single atom, therefore, even small two-body interaction brings about some deviation from 2D kinematics. In the absence of many-body interactions, however, the center-of-mass kinematics of the pair is 2D because the confining potential is close to harmonic.
The whole system, therefore, remains kinematically 2D despite any level of two-body interactions.
For example, the bosonic asymptote of a Fermi-to-Bose crossover is a 2D gas of molecular dimers, while the motion of an atom within a dimer is 3D and may be expanded as a superposition of many eigenstates of potential $m\omega_z^2z^2/2$ in the laboratory reference frame.

In the strongly-interacting regime, at $a_2\sqrt{n_2}\sim1$, the interaction is necessarily of a many-body type. Whether such interaction breaks 2D kinematics is an important and yet unresolved question. Breakdown of 2D kinematics by interactions has been recently addressed in Ref.~\cite{Vale2DCriteria2016}. In particular, it has been found that for $E_F/\hbar\omega_z=0.59$, \textit{i.~e.}, same $E_F/\hbar\omega_z$ as here, the 2D kinematics is broken down for $l_z/a>-0.34$ and obeyed below that threshold. This suggests that the data of Figs.~\ref{fig:ComparePressure}(i--l) are not in the 2D regime, while the data in panels~\ref{fig:ComparePressure}(a--h) are for kinematically 2D gases.
Conclusion of Ref.~\cite{Vale2DCriteria2016} about such numerical criterion of 2D kinematics is opposed below.
This criterion is based on analyzing disc-shaped-cloud expansion after removal of the tight confinement. The expansion law changes sharply in response to increasing atom number and, above the threshold, is faster than predicted by the noninteracting-gas model and a collisional-gas model. This change in dynamics is interpreted as a breakdown of two-dimensionality.
As an alternative to deviation from 2D kinematics one may consider the onset of superfluidity in the 2D system. Onset of superfluidity is plausible for two reasons: (i) a superfluid gas may expand faster than a normal collisional gas and (ii) sharp change is possible at the onset of superfluidity; for example, abrupt change in dynamics has been observed in 3D Fermi gases~\cite{GrimmBreathingMode2004,HydroBreakdown}. Until superfluidity is ruled out, observation of interaction-induced breakdown of 2D kinematics in Ref.~\cite{Vale2DCriteria2016} is controversial and the question of whether many-body interaction breaks 2D kinematics remains open.

The use of the 2D pressure as an indicator of broken 2D kinematics is subtle. Deviation from 2D kinematics would not alter $P(\tilde\rho=0)$ but should result in reduction of $P$ at $\tilde\rho>0$. Determination of whether 2D kinematics is broken requires comparison of measured pressure to a model whose correctness is well established. From comparison to the available 2D models one may see that the pressure does not show significant downshifts relative to these models. This suggests that the departure from 2D is either small or absent.

Comparison with Monte Carlo models, in particular, shows that in all cases, there is a curve which either agrees with or goes below the data. The auxiliary-field quantum Monte Carlo~\cite{Fermi2DExactGS2015} predicts profiles that are below or in agreement with the data in all panels of Fig.~\ref{fig:ComparePressure} except for Fig.~\ref{fig:ComparePressure}(l). This suggest that kinematics is 2D in all cases except for small deviation from 2D for the conditions of Fig.~\ref{fig:ComparePressure}(l).
Comparison with the lattice Monte Carlo~\cite{Fermi2DAbInitioLattice2015} always shows agreement with 2D kinematics.

\section{Conclusion}\label{sec:Conclusion}

Spatial profiles of pressure have been measured in a Fermi gas, whose kinematics is either 2D or close to 2D. The gas is two-component, with \textit{s}-wave attraction, at nearly zero temperature. Planar distribution of the pressure is recovered from a 1D density profile. The method is applicable to both fermionic and bosonic systems.
We show that contribution of finite-temperature effects is small and compare the measured pressure to zero-temperature models within the local density approximation.
For weak interaction, the data are systematically above the zero-temperature Fermi-liquid theory prediction, while the mean-field model with fluctuations gives results, which are the closest to the measured values.
In the regime of strong interactions, in response to the increasing attraction, the pressure shifts below this model reaching lower values calculated within quantum Monte Carlo methods. Comparison to the Monte Carlo results shows that the kinematics of atom pairs may remain 2D despite strong interaction.

\begin{acknowledgements}
The authors acknowledge the financial support by the programs of the Presidium of Russian Academy of Sciences ``Actual problems of low-temperature physics'' and ``Fundamental problems of nonlinear dynamics'' and Russian Foundation for Basic Research (Grants No. 14-22-02080-ofi-m, 14-02-31681-mol\_a, 15-02-08464, 15-42-02638).
\end{acknowledgements}


\begin{thebibliography}{30}%
\makeatletter
\providecommand \@ifxundefined [1]{%
 \@ifx{#1\undefined}
}%
\providecommand \@ifnum [1]{%
 \ifnum #1\expandafter \@firstoftwo
 \else \expandafter \@secondoftwo
 \fi
}%
\providecommand \@ifx [1]{%
 \ifx #1\expandafter \@firstoftwo
 \else \expandafter \@secondoftwo
 \fi
}%
\providecommand \natexlab [1]{#1}%
\providecommand \enquote  [1]{``#1''}%
\providecommand \bibnamefont  [1]{#1}%
\providecommand \bibfnamefont [1]{#1}%
\providecommand \citenamefont [1]{#1}%
\providecommand \href@noop [0]{\@secondoftwo}%
\providecommand \href [0]{\begingroup \@sanitize@url \@href}%
\providecommand \@href[1]{\@@startlink{#1}\@@href}%
\providecommand \@@href[1]{\endgroup#1\@@endlink}%
\providecommand \@sanitize@url [0]{\catcode `\\12\catcode `\$12\catcode
  `\&12\catcode `\#12\catcode `\^12\catcode `\_12\catcode `\%12\relax}%
\providecommand \@@startlink[1]{}%
\providecommand \@@endlink[0]{}%
\providecommand \url  [0]{\begingroup\@sanitize@url \@url }%
\providecommand \@url [1]{\endgroup\@href {#1}{\urlprefix }}%
\providecommand \urlprefix  [0]{URL }%
\providecommand \Eprint [0]{\href }%
\providecommand \doibase [0]{http://dx.doi.org/}%
\providecommand \selectlanguage [0]{\@gobble}%
\providecommand \bibinfo  [0]{\@secondoftwo}%
\providecommand \bibfield  [0]{\@secondoftwo}%
\providecommand \translation [1]{[#1]}%
\providecommand \BibitemOpen [0]{}%
\providecommand \bibitemStop [0]{}%
\providecommand \bibitemNoStop [0]{.\EOS\space}%
\providecommand \EOS [0]{\spacefactor3000\relax}%
\providecommand \BibitemShut  [1]{\csname bibitem#1\endcsname}%
\let\auto@bib@innerbib\@empty
\bibitem [{\citenamefont {Levanyuk}(1959)}]{GinzburgLevanyuk1959eng}%
  \BibitemOpen
  \bibfield  {author} {\bibinfo {author} {\bibfnamefont {A.~P.}\ \bibnamefont
  {Levanyuk}},\ }\bibfield  {title} {\enquote {\bibinfo {title} {Contribution
  to the theory of scattering of light near phase transition points of the
  second kind},}\ }\href@noop {} {\bibfield  {journal} {\bibinfo  {journal}
  {Sov. Phys. JETP}\ }\textbf {\bibinfo {volume} {9}},\ \bibinfo {pages} {571}
  (\bibinfo {year} {1959})}\BibitemShut {NoStop}%
\bibitem [{\citenamefont {Ginzburg}(1960)}]{GinzburgLevanyuk1960eng}%
  \BibitemOpen
  \bibfield  {author} {\bibinfo {author} {\bibfnamefont {V.~L.}\ \bibnamefont
  {Ginzburg}},\ }\href@noop {} {\bibfield  {journal} {\bibinfo  {journal} {Sov.
  Phys. Solid State}\ }\textbf {\bibinfo {volume} {2}},\ \bibinfo {pages}
  {1824} (\bibinfo {year} {1960})}\BibitemShut {NoStop}%
\bibitem [{\citenamefont {Miyake}(1983)}]{Miyake1983}%
  \BibitemOpen
  \bibfield  {author} {\bibinfo {author} {\bibfnamefont {Kazumasa}\
  \bibnamefont {Miyake}},\ }\bibfield  {title} {\enquote {\bibinfo {title}
  {Fermi liquid theory of dilute submonolayer {$^3$He} on thin {$^4$He} {II}
  film: Dimer bound state and {Cooper} pairs},}\ }\href {\doibase
  10.1143/PTP.69.1794} {\bibfield  {journal} {\bibinfo  {journal} {Prog. Theor.
  Phys.}\ }\textbf {\bibinfo {volume} {69}},\ \bibinfo {pages} {1794--1797}
  (\bibinfo {year} {1983})}\BibitemShut {NoStop}%
\bibitem [{\citenamefont {Randeria}\ \emph {et~al.}(1989)\citenamefont
  {Randeria}, \citenamefont {Duan},\ and\ \citenamefont
  {Shieh}}]{Randeria2DCrossover1989prl}%
  \BibitemOpen
  \bibfield  {author} {\bibinfo {author} {\bibfnamefont {Mohit}\ \bibnamefont
  {Randeria}}, \bibinfo {author} {\bibfnamefont {Ji-Min}\ \bibnamefont {Duan}},
  \ and\ \bibinfo {author} {\bibfnamefont {Lih-Yir}\ \bibnamefont {Shieh}},\
  }\bibfield  {title} {\enquote {\bibinfo {title} {Bound states, {Cooper}
  pairing, and {Bose} condensation in two dimensions},}\ }\href {\doibase
  10.1103/PhysRevLett.62.981} {\bibfield  {journal} {\bibinfo  {journal} {Phys.
  Rev. Lett.}\ }\textbf {\bibinfo {volume} {62}},\ \bibinfo {pages} {981--984}
  (\bibinfo {year} {1989})}\BibitemShut {NoStop}%
\bibitem [{\citenamefont {Kagan}(2013)}]{MKaganBook2013}%
  \BibitemOpen
  \bibfield  {author} {\bibinfo {author} {\bibfnamefont {M.~Yu.}\ \bibnamefont
  {Kagan}},\ }\href {\doibase 10.1007/978-94-007-6961-8} {\emph {\bibinfo
  {title} {Modern trends in superconductivity and superfluidity}}},\ \bibinfo
  {series} {Lecture Notes in Physics}, Vol.\ \bibinfo {volume} {874}\ (\bibinfo
   {publisher} {Springer},\ \bibinfo {address} {Dordrecht},\ \bibinfo {year}
  {2013})\BibitemShut {NoStop}%
\bibitem [{\citenamefont {Bertaina}\ and\ \citenamefont
  {Giorgini}(2011)}]{Giorgini2D2011}%
  \BibitemOpen
  \bibfield  {author} {\bibinfo {author} {\bibfnamefont {G.}~\bibnamefont
  {Bertaina}}\ and\ \bibinfo {author} {\bibfnamefont {S.}~\bibnamefont
  {Giorgini}},\ }\bibfield  {title} {\enquote {\bibinfo {title} {{BCS}-{BEC}
  crossover in a two-dimensional {Fermi} gas},}\ }\href {\doibase
  10.1103/PhysRevLett.106.110403} {\bibfield  {journal} {\bibinfo  {journal}
  {Phys. Rev. Lett.}\ }\textbf {\bibinfo {volume} {106}},\ \bibinfo {pages}
  {110403} (\bibinfo {year} {2011})}\BibitemShut {NoStop}%
\bibitem [{\citenamefont {Salasnich}\ and\ \citenamefont
  {Toigo}(2015)}]{SalasnichCompositeBosonsFromFluctuations2015}%
  \BibitemOpen
  \bibfield  {author} {\bibinfo {author} {\bibfnamefont {L.}~\bibnamefont
  {Salasnich}}\ and\ \bibinfo {author} {\bibfnamefont {F.}~\bibnamefont
  {Toigo}},\ }\bibfield  {title} {\enquote {\bibinfo {title} {Composite bosons
  in the two-dimensional {BCS}-{BEC} crossover from {Gaussian} fluctuations},}\
  }\href {\doibase 10.1103/PhysRevA.91.011604} {\bibfield  {journal} {\bibinfo
  {journal} {Phys. Rev. A}\ }\textbf {\bibinfo {volume} {91}},\ \bibinfo
  {pages} {011604} (\bibinfo {year} {2015})}\BibitemShut {NoStop}%
\bibitem [{\citenamefont {He}\ \emph {et~al.}(2015)\citenamefont {He},
  \citenamefont {L\"u}, \citenamefont {Cao}, \citenamefont {Hu},\ and\
  \citenamefont {Liu}}]{Fermi2DMeanFieldPlusFluctuations2015}%
  \BibitemOpen
  \bibfield  {author} {\bibinfo {author} {\bibfnamefont {Lianyi}\ \bibnamefont
  {He}}, \bibinfo {author} {\bibfnamefont {Haifeng}\ \bibnamefont {L\"u}},
  \bibinfo {author} {\bibfnamefont {Gaoqing}\ \bibnamefont {Cao}}, \bibinfo
  {author} {\bibfnamefont {Hui}\ \bibnamefont {Hu}}, \ and\ \bibinfo {author}
  {\bibfnamefont {Xia-Ji}\ \bibnamefont {Liu}},\ }\bibfield  {title} {\enquote
  {\bibinfo {title} {Quantum fluctuations in the {BCS}-{BEC} crossover of
  two-dimensional {Fermi} gases},}\ }\href {\doibase
  10.1103/PhysRevA.92.023620} {\bibfield  {journal} {\bibinfo  {journal} {Phys.
  Rev. A}\ }\textbf {\bibinfo {volume} {92}},\ \bibinfo {pages} {023620}
  (\bibinfo {year} {2015})}\BibitemShut {NoStop}%
\bibitem [{\citenamefont {Bloom}(1975)}]{Bloom1975}%
  \BibitemOpen
  \bibfield  {author} {\bibinfo {author} {\bibfnamefont {Paul}\ \bibnamefont
  {Bloom}},\ }\bibfield  {title} {\enquote {\bibinfo {title} {Two-dimensional
  {Fermi} gas},}\ }\href {\doibase 10.1103/PhysRevB.12.125} {\bibfield
  {journal} {\bibinfo  {journal} {Phys. Rev. B}\ }\textbf {\bibinfo {volume}
  {12}},\ \bibinfo {pages} {125--129} (\bibinfo {year} {1975})}\BibitemShut
  {NoStop}%
\bibitem [{\citenamefont {Engelbrecht}\ \emph {et~al.}(1992)\citenamefont
  {Engelbrecht}, \citenamefont {Randeria},\ and\ \citenamefont
  {Zhang}}]{FermiLiquid2D1992}%
  \BibitemOpen
  \bibfield  {author} {\bibinfo {author} {\bibfnamefont {Jan~R.}\ \bibnamefont
  {Engelbrecht}}, \bibinfo {author} {\bibfnamefont {Mohit}\ \bibnamefont
  {Randeria}}, \ and\ \bibinfo {author} {\bibfnamefont {Lizeng}\ \bibnamefont
  {Zhang}},\ }\bibfield  {title} {\enquote {\bibinfo {title} {Landau \textit{f}
  function for the dilute {Fermi} gas in two dimensions},}\ }\href {\doibase
  10.1103/PhysRevB.45.10135} {\bibfield  {journal} {\bibinfo  {journal} {Phys.
  Rev. B}\ }\textbf {\bibinfo {volume} {45}},\ \bibinfo {pages} {10135--10138}
  (\bibinfo {year} {1992})}\BibitemShut {NoStop}%
\bibitem [{\citenamefont {Galea}\ \emph {et~al.}(2016)\citenamefont {Galea},
  \citenamefont {Dawkins}, \citenamefont {Gandolfi},\ and\ \citenamefont
  {Gezerlis}}]{DiffusionMC2015}%
  \BibitemOpen
  \bibfield  {author} {\bibinfo {author} {\bibfnamefont {Alexander}\
  \bibnamefont {Galea}}, \bibinfo {author} {\bibfnamefont {Hillary}\
  \bibnamefont {Dawkins}}, \bibinfo {author} {\bibfnamefont {Stefano}\
  \bibnamefont {Gandolfi}}, \ and\ \bibinfo {author} {\bibfnamefont
  {Alexandros}\ \bibnamefont {Gezerlis}},\ }\bibfield  {title} {\enquote
  {\bibinfo {title} {Diffusion {Monte} {Carlo} study of strongly interacting
  two-dimensional {Fermi} gases},}\ }\href {\doibase
  10.1103/PhysRevA.93.023602} {\bibfield  {journal} {\bibinfo  {journal} {Phys.
  Rev. A}\ }\textbf {\bibinfo {volume} {93}},\ \bibinfo {pages} {023602}
  (\bibinfo {year} {2016})}\BibitemShut {NoStop}%
\bibitem [{\citenamefont {Bauer}\ \emph {et~al.}(2014)\citenamefont {Bauer},
  \citenamefont {Parish},\ and\ \citenamefont
  {Enss}}]{Fermi2DEOSandPressureParish2014}%
  \BibitemOpen
  \bibfield  {author} {\bibinfo {author} {\bibfnamefont {Marianne}\
  \bibnamefont {Bauer}}, \bibinfo {author} {\bibfnamefont {Meera~M.}\
  \bibnamefont {Parish}}, \ and\ \bibinfo {author} {\bibfnamefont {Tilman}\
  \bibnamefont {Enss}},\ }\bibfield  {title} {\enquote {\bibinfo {title}
  {Universal equation of state and pseudogap in the two-dimensional {Fermi}
  gas},}\ }\href {\doibase 10.1103/PhysRevLett.112.135302} {\bibfield
  {journal} {\bibinfo  {journal} {Phys. Rev. Lett.}\ }\textbf {\bibinfo
  {volume} {112}},\ \bibinfo {pages} {135302} (\bibinfo {year}
  {2014})}\BibitemShut {NoStop}%
\bibitem [{\citenamefont {Anderson}\ and\ \citenamefont
  {Drut}(2015)}]{Fermi2DAbInitioLattice2015}%
  \BibitemOpen
  \bibfield  {author} {\bibinfo {author} {\bibfnamefont {E.~R.}\ \bibnamefont
  {Anderson}}\ and\ \bibinfo {author} {\bibfnamefont {J.~E.}\ \bibnamefont
  {Drut}},\ }\bibfield  {title} {\enquote {\bibinfo {title} {Pressure,
  compressibility, and contact of the two-dimensional attractive {Fermi}
  gas},}\ }\href {\doibase 10.1103/PhysRevLett.115.115301} {\bibfield
  {journal} {\bibinfo  {journal} {Phys. Rev. Lett.}\ }\textbf {\bibinfo
  {volume} {115}},\ \bibinfo {pages} {115301} (\bibinfo {year}
  {2015})}\BibitemShut {NoStop}%
\bibitem [{\citenamefont {Shi}\ \emph {et~al.}(2015)\citenamefont {Shi},
  \citenamefont {Chiesa},\ and\ \citenamefont {Zhang}}]{Fermi2DExactGS2015}%
  \BibitemOpen
  \bibfield  {author} {\bibinfo {author} {\bibfnamefont {Hao}\ \bibnamefont
  {Shi}}, \bibinfo {author} {\bibfnamefont {Simone}\ \bibnamefont {Chiesa}}, \
  and\ \bibinfo {author} {\bibfnamefont {Shiwei}\ \bibnamefont {Zhang}},\
  }\bibfield  {title} {\enquote {\bibinfo {title} {Ground-state properties of
  strongly interacting {Fermi} gases in two dimensions},}\ }\href {\doibase
  10.1103/PhysRevA.92.033603} {\bibfield  {journal} {\bibinfo  {journal} {Phys.
  Rev. A}\ }\textbf {\bibinfo {volume} {92}},\ \bibinfo {pages} {033603}
  (\bibinfo {year} {2015})}\BibitemShut {NoStop}%
\bibitem [{\citenamefont {Martiyanov}\ \emph {et~al.}(2010)\citenamefont
  {Martiyanov}, \citenamefont {Makhalov},\ and\ \citenamefont
  {Turlapov}}]{Fermi2D}%
  \BibitemOpen
  \bibfield  {author} {\bibinfo {author} {\bibfnamefont {Kirill}\ \bibnamefont
  {Martiyanov}}, \bibinfo {author} {\bibfnamefont {Vasiliy}\ \bibnamefont
  {Makhalov}}, \ and\ \bibinfo {author} {\bibfnamefont {Andrey}\ \bibnamefont
  {Turlapov}},\ }\bibfield  {title} {\enquote {\bibinfo {title} {Observation of
  a two-dimensional {Fermi} gas of atoms},}\ }\href {\doibase
  10.1103/PhysRevLett.105.030404} {\bibfield  {journal} {\bibinfo  {journal}
  {Phys. Rev. Lett.}\ }\textbf {\bibinfo {volume} {105}},\ \bibinfo {pages}
  {030404} (\bibinfo {year} {2010})}\BibitemShut {NoStop}%
\bibitem [{\citenamefont {Makhalov}\ \emph {et~al.}(2014)\citenamefont
  {Makhalov}, \citenamefont {Martiyanov},\ and\ \citenamefont
  {Turlapov}}]{FermiBose2DCrossover}%
  \BibitemOpen
  \bibfield  {author} {\bibinfo {author} {\bibfnamefont {Vasiliy}\ \bibnamefont
  {Makhalov}}, \bibinfo {author} {\bibfnamefont {Kirill}\ \bibnamefont
  {Martiyanov}}, \ and\ \bibinfo {author} {\bibfnamefont {Andrey}\ \bibnamefont
  {Turlapov}},\ }\bibfield  {title} {\enquote {\bibinfo {title} {Ground-state
  pressure of quasi-{2D} {Fermi} and {Bose} gases},}\ }\href {\doibase
  10.1103/PhysRevLett.112.045301} {\bibfield  {journal} {\bibinfo  {journal}
  {Phys. Rev. Lett.}\ }\textbf {\bibinfo {volume} {112}},\ \bibinfo {pages}
  {045301} (\bibinfo {year} {2014})}\BibitemShut {NoStop}%
\bibitem [{\citenamefont {Boettcher}\ \emph {et~al.}(2016)\citenamefont
  {Boettcher}, \citenamefont {Bayha}, \citenamefont {Kedar}, \citenamefont
  {Murthy}, \citenamefont {Neidig}, \citenamefont {Ries}, \citenamefont {Wenz},
  \citenamefont {Z\"urn}, \citenamefont {Jochim},\ and\ \citenamefont
  {Enss}}]{Jochim2DThermodynamics2015}%
  \BibitemOpen
  \bibfield  {author} {\bibinfo {author} {\bibfnamefont {I.}~\bibnamefont
  {Boettcher}}, \bibinfo {author} {\bibfnamefont {L.}~\bibnamefont {Bayha}},
  \bibinfo {author} {\bibfnamefont {D.}~\bibnamefont {Kedar}}, \bibinfo
  {author} {\bibfnamefont {P.~A.}\ \bibnamefont {Murthy}}, \bibinfo {author}
  {\bibfnamefont {M.}~\bibnamefont {Neidig}}, \bibinfo {author} {\bibfnamefont
  {M.~G.}\ \bibnamefont {Ries}}, \bibinfo {author} {\bibfnamefont {A.~N.}\
  \bibnamefont {Wenz}}, \bibinfo {author} {\bibfnamefont {G.}~\bibnamefont
  {Z\"urn}}, \bibinfo {author} {\bibfnamefont {S.}~\bibnamefont {Jochim}}, \
  and\ \bibinfo {author} {\bibfnamefont {T.}~\bibnamefont {Enss}},\ }\bibfield
  {title} {\enquote {\bibinfo {title} {Equation of state of ultracold fermions
  in the {2D} {BEC}-{BCS} crossover region},}\ }\href {\doibase
  10.1103/PhysRevLett.116.045303} {\bibfield  {journal} {\bibinfo  {journal}
  {Phys. Rev. Lett.}\ }\textbf {\bibinfo {volume} {116}},\ \bibinfo {pages}
  {045303} (\bibinfo {year} {2016})}\BibitemShut {NoStop}%
\bibitem [{\citenamefont {Fenech}\ \emph {et~al.}(2016)\citenamefont {Fenech},
  \citenamefont {Dyke}, \citenamefont {Peppler}, \citenamefont {Lingham},
  \citenamefont {Hoinka}, \citenamefont {Hu},\ and\ \citenamefont
  {Vale}}]{Vale2DThermodynamics2015}%
  \BibitemOpen
  \bibfield  {author} {\bibinfo {author} {\bibfnamefont {K.}~\bibnamefont
  {Fenech}}, \bibinfo {author} {\bibfnamefont {P.}~\bibnamefont {Dyke}},
  \bibinfo {author} {\bibfnamefont {T.}~\bibnamefont {Peppler}}, \bibinfo
  {author} {\bibfnamefont {M.~G.}\ \bibnamefont {Lingham}}, \bibinfo {author}
  {\bibfnamefont {S.}~\bibnamefont {Hoinka}}, \bibinfo {author} {\bibfnamefont
  {H.}~\bibnamefont {Hu}}, \ and\ \bibinfo {author} {\bibfnamefont {C.~J.}\
  \bibnamefont {Vale}},\ }\bibfield  {title} {\enquote {\bibinfo {title}
  {Thermodynamics of an attractive {2D} {Fermi} gas},}\ }\href {\doibase
  10.1103/PhysRevLett.116.045302} {\bibfield  {journal} {\bibinfo  {journal}
  {Phys. Rev. Lett.}\ }\textbf {\bibinfo {volume} {116}},\ \bibinfo {pages}
  {045302} (\bibinfo {year} {2016})}\BibitemShut {NoStop}%
\bibitem [{\citenamefont {Thomas}\ \emph {et~al.}(2005)\citenamefont {Thomas},
  \citenamefont {Kinast},\ and\ \citenamefont {Turlapov}}]{ThomasUniversal}%
  \BibitemOpen
  \bibfield  {author} {\bibinfo {author} {\bibfnamefont {J.~E.}\ \bibnamefont
  {Thomas}}, \bibinfo {author} {\bibfnamefont {J.}~\bibnamefont {Kinast}}, \
  and\ \bibinfo {author} {\bibfnamefont {A.}~\bibnamefont {Turlapov}},\
  }\bibfield  {title} {\enquote {\bibinfo {title} {Virial theorem and
  universality in a unitary {Fermi} gas},}\ }\href {\doibase
  10.1103/PhysRevLett.95.120402} {\bibfield  {journal} {\bibinfo  {journal}
  {Phys. Rev. Lett.}\ }\textbf {\bibinfo {volume} {95}},\ \bibinfo {pages}
  {120402} (\bibinfo {year} {2005})}\BibitemShut {NoStop}%
\bibitem [{\citenamefont {Ho}\ and\ \citenamefont
  {Zhou}(2010)}]{HoLocalMeas2009}%
  \BibitemOpen
  \bibfield  {author} {\bibinfo {author} {\bibfnamefont {Tin-Lun}\ \bibnamefont
  {Ho}}\ and\ \bibinfo {author} {\bibfnamefont {Qi}~\bibnamefont {Zhou}},\
  }\bibfield  {title} {\enquote {\bibinfo {title} {Obtaining the phase diagram
  and thermodynamic quantities of bulk systems from the densities of trapped
  gases},}\ }\href {\doibase 10.1038/nphys1477} {\bibfield  {journal} {\bibinfo
   {journal} {Nature Phys.}\ }\textbf {\bibinfo {volume} {6}},\ \bibinfo
  {pages} {131} (\bibinfo {year} {2010})}\BibitemShut {NoStop}%
\bibitem [{\citenamefont {Makhalov}\ \emph {et~al.}(2015)\citenamefont
  {Makhalov}, \citenamefont {Martiyanov}, \citenamefont {Barmashova},\ and\
  \citenamefont {Turlapov}}]{Parametric2015}%
  \BibitemOpen
  \bibfield  {author} {\bibinfo {author} {\bibfnamefont {Vasiliy}\ \bibnamefont
  {Makhalov}}, \bibinfo {author} {\bibfnamefont {Kirill}\ \bibnamefont
  {Martiyanov}}, \bibinfo {author} {\bibfnamefont {Tatiana}\ \bibnamefont
  {Barmashova}}, \ and\ \bibinfo {author} {\bibfnamefont {Andrey}\ \bibnamefont
  {Turlapov}},\ }\bibfield  {title} {\enquote {\bibinfo {title} {Precision
  measurement of a trapping potential for an ultracold gas},}\ }\href {\doibase
  j.physleta.2014.10.049} {\bibfield  {journal} {\bibinfo  {journal} {Physics
  Letters A}\ }\textbf {\bibinfo {volume} {379}},\ \bibinfo {pages} {327--332}
  (\bibinfo {year} {2015})},\ \bibinfo {note} {published online 29 Nov.
  2014}\BibitemShut {NoStop}%
\bibitem [{\citenamefont {Chin}\ \emph {et~al.}(2010)\citenamefont {Chin},
  \citenamefont {Grimm}, \citenamefont {Julienne},\ and\ \citenamefont
  {Tiesinga}}]{FeshbachReview2010}%
  \BibitemOpen
  \bibfield  {author} {\bibinfo {author} {\bibfnamefont {Cheng}\ \bibnamefont
  {Chin}}, \bibinfo {author} {\bibfnamefont {Rudolf}\ \bibnamefont {Grimm}},
  \bibinfo {author} {\bibfnamefont {Paul}\ \bibnamefont {Julienne}}, \ and\
  \bibinfo {author} {\bibfnamefont {Eite}\ \bibnamefont {Tiesinga}},\
  }\bibfield  {title} {\enquote {\bibinfo {title} {Feshbach resonances in
  ultracold gases},}\ }\href {\doibase 10.1103/RevModPhys.82.1225} {\bibfield
  {journal} {\bibinfo  {journal} {Rev. Mod. Phys.}\ }\textbf {\bibinfo {volume}
  {82}},\ \bibinfo {pages} {1225--1286} (\bibinfo {year} {2010})}\BibitemShut
  {NoStop}%
\bibitem [{\citenamefont {Z\"urn}\ \emph {et~al.}(2013)\citenamefont {Z\"urn},
  \citenamefont {Lompe}, \citenamefont {Wenz}, \citenamefont {Jochim},
  \citenamefont {Julienne},\ and\ \citenamefont
  {Hutson}}]{JochimNewLiFeshbach2013}%
  \BibitemOpen
  \bibfield  {author} {\bibinfo {author} {\bibfnamefont {G.}~\bibnamefont
  {Z\"urn}}, \bibinfo {author} {\bibfnamefont {T.}~\bibnamefont {Lompe}},
  \bibinfo {author} {\bibfnamefont {A.~N.}\ \bibnamefont {Wenz}}, \bibinfo
  {author} {\bibfnamefont {S.}~\bibnamefont {Jochim}}, \bibinfo {author}
  {\bibfnamefont {P.~S.}\ \bibnamefont {Julienne}}, \ and\ \bibinfo {author}
  {\bibfnamefont {J.~M.}\ \bibnamefont {Hutson}},\ }\bibfield  {title}
  {\enquote {\bibinfo {title} {Precise characterization of $^{6}\mathrm{Li}$
  {Feshbach} resonances using trap-sideband-resolved rf spectroscopy of weakly
  bound molecules},}\ }\href {\doibase 10.1103/PhysRevLett.110.135301}
  {\bibfield  {journal} {\bibinfo  {journal} {Phys. Rev. Lett.}\ }\textbf
  {\bibinfo {volume} {110}},\ \bibinfo {pages} {135301} (\bibinfo {year}
  {2013})}\BibitemShut {NoStop}%
\bibitem [{\citenamefont {Petrov}\ and\ \citenamefont
  {Shlyapnikov}(2001)}]{Shlyapnikov2DScattering2001}%
  \BibitemOpen
  \bibfield  {author} {\bibinfo {author} {\bibfnamefont {D.~S.}\ \bibnamefont
  {Petrov}}\ and\ \bibinfo {author} {\bibfnamefont {G.~V.}\ \bibnamefont
  {Shlyapnikov}},\ }\bibfield  {title} {\enquote {\bibinfo {title} {Interatomic
  collisions in a tightly confined {Bose} gas},}\ }\href {\doibase
  10.1103/PhysRevA.64.012706} {\bibfield  {journal} {\bibinfo  {journal} {Phys.
  Rev. A}\ }\textbf {\bibinfo {volume} {64}},\ \bibinfo {pages} {012706}
  (\bibinfo {year} {2001})}\BibitemShut {NoStop}%
\bibitem [{\citenamefont {Barmashova}\ \emph {et~al.}(2016)\citenamefont
  {Barmashova}, \citenamefont {Martiyanov}, \citenamefont {Makhalov},\ and\
  \citenamefont {Turlapov}}]{UFN2016eng}%
  \BibitemOpen
  \bibfield  {author} {\bibinfo {author} {\bibfnamefont {T.~V.}\ \bibnamefont
  {Barmashova}}, \bibinfo {author} {\bibfnamefont {K.~A.}\ \bibnamefont
  {Martiyanov}}, \bibinfo {author} {\bibfnamefont {V.~B.}\ \bibnamefont
  {Makhalov}}, \ and\ \bibinfo {author} {\bibfnamefont {A.~V.}\ \bibnamefont
  {Turlapov}},\ }\bibfield  {title} {\enquote {\bibinfo {title} {Fermi liquid
  to {Bose} condensate crossover in a two-dimensional ultracold gas
  experiment},}\ }\href {\doibase 10.3367/UFNe.0186.201602i.0183} {\bibfield
  {journal} {\bibinfo  {journal} {Phys. Usp.}\ }\textbf {\bibinfo {volume}
  {59}},\ \bibinfo {pages} {174--183} (\bibinfo {year} {2016})}\BibitemShut
  {NoStop}%
\bibitem [{\citenamefont {Bloch}\ \emph {et~al.}(2008)\citenamefont {Bloch},
  \citenamefont {Dalibard},\ and\ \citenamefont
  {Zwerger}}]{BlochLowDReview2008}%
  \BibitemOpen
  \bibfield  {author} {\bibinfo {author} {\bibfnamefont {Immanuel}\
  \bibnamefont {Bloch}}, \bibinfo {author} {\bibfnamefont {Jean}\ \bibnamefont
  {Dalibard}}, \ and\ \bibinfo {author} {\bibfnamefont {Wilhelm}\ \bibnamefont
  {Zwerger}},\ }\bibfield  {title} {\enquote {\bibinfo {title} {Many-body
  physics with ultracold gases},}\ }\href {\doibase 10.1103/RevModPhys.80.885}
  {\bibfield  {journal} {\bibinfo  {journal} {Rev. Mod. Phys.}\ }\textbf
  {\bibinfo {volume} {80}},\ \bibinfo {pages} {885--964} (\bibinfo {year}
  {2008})}\BibitemShut {NoStop}%
\bibitem [{\citenamefont {Petrov}\ \emph {et~al.}(2003)\citenamefont {Petrov},
  \citenamefont {Baranov},\ and\ \citenamefont
  {Shlyapnikov}}]{Shlyapnikov2DFermi2003}%
  \BibitemOpen
  \bibfield  {author} {\bibinfo {author} {\bibfnamefont {D.~S.}\ \bibnamefont
  {Petrov}}, \bibinfo {author} {\bibfnamefont {M.~A.}\ \bibnamefont {Baranov}},
  \ and\ \bibinfo {author} {\bibfnamefont {G.~V.}\ \bibnamefont
  {Shlyapnikov}},\ }\bibfield  {title} {\enquote {\bibinfo {title} {Superfluid
  transition in quasi-two-dimensional {Fermi} gases},}\ }\href {\doibase
  10.1103/PhysRevA.67.031601} {\bibfield  {journal} {\bibinfo  {journal} {Phys.
  Rev. A}\ }\textbf {\bibinfo {volume} {67}},\ \bibinfo {pages} {031601}
  (\bibinfo {year} {2003})}\BibitemShut {NoStop}%
\bibitem [{\citenamefont {Dyke}\ \emph {et~al.}(2016)\citenamefont {Dyke},
  \citenamefont {Fenech}, \citenamefont {Peppler}, \citenamefont {Lingham},
  \citenamefont {Hoinka}, \citenamefont {Zhang}, \citenamefont {Peng},
  \citenamefont {Mulkerin}, \citenamefont {Hu}, \citenamefont {Liu},\ and\
  \citenamefont {Vale}}]{Vale2DCriteria2016}%
  \BibitemOpen
  \bibfield  {author} {\bibinfo {author} {\bibfnamefont {P.}~\bibnamefont
  {Dyke}}, \bibinfo {author} {\bibfnamefont {K.}~\bibnamefont {Fenech}},
  \bibinfo {author} {\bibfnamefont {T.}~\bibnamefont {Peppler}}, \bibinfo
  {author} {\bibfnamefont {M.~G.}\ \bibnamefont {Lingham}}, \bibinfo {author}
  {\bibfnamefont {S.}~\bibnamefont {Hoinka}}, \bibinfo {author} {\bibfnamefont
  {W.}~\bibnamefont {Zhang}}, \bibinfo {author} {\bibfnamefont {S.-G.}\
  \bibnamefont {Peng}}, \bibinfo {author} {\bibfnamefont {B.}~\bibnamefont
  {Mulkerin}}, \bibinfo {author} {\bibfnamefont {H.}~\bibnamefont {Hu}},
  \bibinfo {author} {\bibfnamefont {X.-J.}\ \bibnamefont {Liu}}, \ and\
  \bibinfo {author} {\bibfnamefont {C.~J.}\ \bibnamefont {Vale}},\ }\bibfield
  {title} {\enquote {\bibinfo {title} {Criteria for two-dimensional kinematics
  in an interacting {Fermi} gas},}\ }\href {\doibase
  10.1103/PhysRevA.93.011603} {\bibfield  {journal} {\bibinfo  {journal} {Phys.
  Rev. A}\ }\textbf {\bibinfo {volume} {93}},\ \bibinfo {pages} {011603}
  (\bibinfo {year} {2016})}\BibitemShut {NoStop}%
\bibitem [{\citenamefont {Bartenstein}\ \emph {et~al.}(2004)\citenamefont
  {Bartenstein}, \citenamefont {Altmeyer}, \citenamefont {Riedl}, \citenamefont
  {Jochim}, \citenamefont {Chin}, \citenamefont {Denschlag},\ and\
  \citenamefont {Grimm}}]{GrimmBreathingMode2004}%
  \BibitemOpen
  \bibfield  {author} {\bibinfo {author} {\bibfnamefont {M.}~\bibnamefont
  {Bartenstein}}, \bibinfo {author} {\bibfnamefont {A.}~\bibnamefont
  {Altmeyer}}, \bibinfo {author} {\bibfnamefont {S.}~\bibnamefont {Riedl}},
  \bibinfo {author} {\bibfnamefont {S.}~\bibnamefont {Jochim}}, \bibinfo
  {author} {\bibfnamefont {C.}~\bibnamefont {Chin}}, \bibinfo {author}
  {\bibfnamefont {J.~Hecker}\ \bibnamefont {Denschlag}}, \ and\ \bibinfo
  {author} {\bibfnamefont {R.}~\bibnamefont {Grimm}},\ }\bibfield  {title}
  {\enquote {\bibinfo {title} {Collective excitations of a degenerate gas at
  the {BEC-BCS} crossover},}\ }\href {\doibase 10.1103/PhysRevLett.92.203201}
  {\bibfield  {journal} {\bibinfo  {journal} {Phys. Rev. Lett.}\ }\textbf
  {\bibinfo {volume} {92}},\ \bibinfo {pages} {203201} (\bibinfo {year}
  {2004})}\BibitemShut {NoStop}%
\bibitem [{\citenamefont {Kinast}\ \emph {et~al.}(2004)\citenamefont {Kinast},
  \citenamefont {Turlapov},\ and\ \citenamefont {Thomas}}]{HydroBreakdown}%
  \BibitemOpen
  \bibfield  {author} {\bibinfo {author} {\bibfnamefont {J.}~\bibnamefont
  {Kinast}}, \bibinfo {author} {\bibfnamefont {A.}~\bibnamefont {Turlapov}}, \
  and\ \bibinfo {author} {\bibfnamefont {J.~E.}\ \bibnamefont {Thomas}},\
  }\bibfield  {title} {\enquote {\bibinfo {title} {Breakdown of hydrodynamics
  in the radial breathing mode of a strongly interacting {Fermi} gas},}\ }\href
  {\doibase 10.1103/PhysRevA.70.051401} {\bibfield  {journal} {\bibinfo
  {journal} {Phys. Rev. A}\ }\textbf {\bibinfo {volume} {70}},\ \bibinfo
  {pages} {051401} (\bibinfo {year} {2004})}\BibitemShut {NoStop}%
\end{thebibliography}
%

\end{document}